\documentclass[submit]{aa}

\usepackage[normalem]{ulem}
\usepackage{graphicx}
\usepackage{amsmath}
\usepackage{amssymb}
\usepackage{placeins}

\usepackage{txfonts}
\usepackage{color}
\usepackage[colorlinks,linkcolor=magenta,anchorcolor=blue,citecolor=blue]{hyperref}
\begin{document} 

   \title{Double radio relics and radio halo in the high redshift galaxy cluster El Gordo with the Upgraded GMRT}

   \subtitle{}

\author{R. Kale
          \inst{1}\fnmsep\thanks{ruta@ncra.tifr.res.in}
          \and
          A. Botteon\inst{2}
          \and 
          D. Eckert \inst{3}
          \and
          R. Santra \inst{1,2}
          \and
          G. Brunetti \inst{2}
          \and 
          T. Venturi \inst{2}
          \and
          R. Cassano \inst{2}
          \and
          D. Dallacasa \inst{4}
          }
\institute{National Centre for Radio Astrophysics, Tata Institute of Fundamental Research, S. P. Pune University Campus, Ganeshkhind, Pune 411007, India
\and
INAF - IRA, Via Gobetti 101, I-40129 Bologna, Italy
\and
Department of Astronomy, University of Geneva, Ch. d’Ecogia 16, CH-1290 Versoix, Switzerland
\and
Dipartimento di Fisica e Astronomia, Università di Bologna, via P. Gobetti 93/2, 40129, Bologna, Italy
}



  \abstract
  {Diffuse synchrotron radio sources associated with the intra-cluster medium of galaxy clusters are of special interest at high redshifts to understand the magnetization and particle acceleration mechanisms.}
  {El Gordo is the most massive galaxy cluster at high redshift ($z = 0.87$), hosts a radio halo and a double radio relic system. We aim to understand the role of turbulence in the origin of the diffuse radio emission by combining radio and X-ray observations.}
  {We observed El Gordo with the Upgraded GMRT at 0.3–1.45 GHz and obtained the integrated spectra, spatially resolved spectral map, and scaling relations between radio and X-ray surface brightness. We constructed a density fluctuation power spectrum for the central 1 Mpc region using \emph{Chandra} data.}
   {The radio halo and the double relics are detected at all the bands and, in addition, we detect an extension to the eastern relic. The radio halo has a spectral index of $-1.0\pm0.3$ 
   with a possible steepening beyond 1.45 GHz. All the 
   relics have spectral indices of $-1.4$ except the extension of the east relic which has $-2.1\pm0.4$. The
   radio and X-ray surface brightness point-to-point analysis at bands 3 and 4 show slopes of $0.60\pm0.12$ and $0.76\pm0.12$, respectively. The spectral index and X-ray surface brightness show an anti-correlation.
   The density fluctuations peak at $\sim 700$ kpc with an amplitude of $(\delta \rho/\rho) =0.15\pm0.02$. We derive the 3D turbulent Mach number of $\sim$ 0.6 from the gas density fluctuations power spectrum, assuming all the fluctuations are attributed to turbulence.}
   {The derived properties of El Gordo are in line with the low redshift clusters indicating that fast magnetic amplification proposed in high redshift clusters is at work in El Gordo as well. We have discussed the consistency of the obtained results with the turbulent re-acceleration which might be representative of high redshift merging clusters.}



   \maketitle


\section{Introduction}
 \label{sec:intro}

A fraction of massive ($M\gtrsim5\times10^{14}\,\, \rm{M}_{\odot}$), X-ray luminous ($L_X\gtrsim8\times10^{44} \,\,\rm{erg}\,\,\rm{s}^{-1}$) clusters that are merging, are known to host diffuse radio emission \citep[e.g.,][]{ven08,cas10,kal15,cuc15, cuc21}. These diffuse sources provide direct evidence for the presence of relativistic electrons and magnetic fields in the intra-cluster medium (ICM) \citep[see][for reviews]{bru14, wee19}. The extended megaparsec-scale radio sources located at cluster centers are termed as radio halos and those with arc-like morphology and located at cluster peripheries are termed as radio relics. Shocks were proposed to generate the radio relic (RR) emission through the mechanism of Diffusive shock acceleration (DSA) \citep[e.g.,][]{ens98, hoe07}. Observational evidence such as spectral index gradients, polarization, and detection of shocks in X-rays has established the role of shocks in the generation of relics \citep[e.g.,][]{gia08,wee10,kal12, 2018ApJ...852...65R, 2022A&A...659A.146D}. However, the major problem with DSA is the observed high luminosity of relics requiring high acceleration efficiencies for accelerating electrons from the thermal pool \citep{kan05,kan11,2020A&A...634A..64B}.

Radio halos (RH) need an {\it in-situ} mechanism of generation owing to the very long diffusion times ($\gtrsim $Gyr) of electrons in the ICM relative to their short radiative lifetimes ($\sim 0.01 - 0.1$ Gyr). Hadronic collisions in the ICM produce secondary relativistic electrons which may be contributing to the radio halos \citep[e.g.,][]{den80,dol00} but are unable to explain the radio halos in the absence of detection of gamma rays and if they are the only sources of relativistic electrons considered \citep{bru17,2021A&A...648A..60A, 2024arXiv240509384O}. Re-acceleration of electrons by MHD turbulence generated by mergers in the ICM has been proposed for the generation of radio halos \citep[][]{bru01,pet01,bru07,miniati15, Fujita_2015,bru17}. Although these mechanisms naturally explain the connection between radio halos and cluster dynamical states \citep[e.g.,][]{cas10,kal15}, several questions about the efficiency of the acceleration mechanism and the details of the plasma processes involved remain open \citep[e.g.,][]{bru14}. Differences in the acceleration mechanisms and in the role played by secondary electrons can lead to observational differences in the spectral properties of radio halos \citep[e.g.,][]{bru11,bru&laz11,pin17,bru17}. These observed spectral features are due to the competition between the re-acceleration and the energy losses, which become more important in the high redshifts \citep[e.g.,][]{2021A&A...646A.135R}.

While the majority of detected radio halos have been at low redshifts (z \textless  0.4) \citep[e.g.,][]{ven08,bon14,kal15,cuc21, 2021PASA...38...10D, 2024PASA...41...26D}, recently with sensitive observations using the LOw Frequency ARray (LOFAR), and MeerKAT, a large number of radio halos at redshifts beyond 0.6 have been uncovered \citep[e.g.,][]{cas19,2021NatAs...5..268D,2021A&A...648A..11O,2022A&A...660A..78B,2021A&A...650A.153D,2024arXiv240403944S}. The newly uncovered population of high redshift radio halos have radio powers similar to those of the nearby radio halos and indicate that the magnetic fields of the same order must be present at the higher redshifts. The spectral indices of these clusters show that half of them have spectral indices\footnote{$S_{\nu}\propto \nu^{\alpha}$, where $S_\nu$ is the flux density at frequency $\nu$ and $\alpha$ is the spectral index.} steeper than $-$1.4 which is expected due to higher IC losses, as a consequence, the electron lifetime is much shorter. \citep{2021A&A...654A.166D, 2024arXiv240403944S}.

The first high redshift cluster discovered to host radio relic and radio halo system was the cluster ACT-CL J0102-4915, dubbed as El Gordo \citep{2012ApJ...748....7M}. El Gordo (SPT-CL J0102-4915) is a galaxy cluster at a redshift of 0.870 with an average temperature (kT) of 
$14.5\pm0.1$ keV and a mass, $M_{200} = (2.16\pm0.32)\times 10^{15}$ h$_{70}^{-1}$ M$_{\odot}$ 
\citep[e.g.,][]{2014ApJ...785...20J}. It is the most massive cluster known beyond the redshift of 0.6 and was discovered using the Sunyaev-Zeldovich effect \citep[SZE][]{1972CoASP...4..173S} with the Atacama Cosmology Telescope \citep{2011ApJ...737...61M} and confirmed to be at a redshift of 0.870 using optical observations \citep{2010ApJ...723.1523M}. The velocity dispersion of the cluster is $1321\pm106$ km s$^{-1}$ \citep{2013ApJ...772...25S}.

The signatures of a merger in El Gordo were first seen in the distinctive wake in the {\it Chandra} X-ray surface brightness \citep{2012ApJ...748....7M} that appear like two tails \citep{2015ApJ...800...37M}. Simulations studies, that have attempted to reproduce the properties of El Gordo infer high initial relative velocities for on and off-axis mergers \citep{2015ApJ...800...37M,2015ApJ...813..129Z} or a highly disturbed initial cluster \citep{2014MNRAS.438.1971D}. 

El Gordo hosts a bright radio relic (northwest relic, NW-relic hereafter) that was discovered using the 610 MHz Giant Metrewave Radio Telescope (GMRT) and 2.1 GHz Australia Telescope Compact Array (ATCA) observations \citep[][hereafter L14]{2014ApJ...786...49L}. The same work also found evidence for the presence of a faint radio halo and a counter relic. The NW-relic has integrated polarized flux fraction of $\sim 33\%$ (at 2.1 GHz) supporting the scenario of Fermi acceleration at a shock during the cluster merger. A subsequent GMRT 610 MHz study confirmed the radio halo and indicated that it followed the northern tail of the X-ray emission \citep{2016MNRAS.463.1534B}. In the same work, deep X-ray observations with \emph{Chandra} revealed a $M\gtrsim3$ shock at the location of the bright relic, and a shock acceleration of electrons from the thermal pool has been proposed to be a viable origin. Recently, the magnetic field has been mapped by \cite{2024NatCo..15.1006H} utilizing the Synchrotron Intensity Gradient (SIG) using observations from MeerKAT at 1.28 GHz. 

In this paper, we report observations of El Gordo with the Upgraded GMRT (uGMRT) in the frequency range of 300 - 1450 MHz. The paper is organized as follows. The radio and X-ray observations and data analysis are described in Sec.~\ref{ugobs}, and Sec.~\ref{xrayobs}, respectively. The radio images, integrated spectra, spectral index map, and the radio and X-ray comparative analysis are presented in Sec.~\ref{results}. The results are discussed in Sec.~\ref{discussion}, with comparing the theoretical models. We use $\Lambda$CDM cosmology with $\rm H_0 = 70 $ km~s$^{-1}$~Mpc$^{-1}$ with $\Omega_{\rm M} = 0.27$ and $\Omega_{\Lambda}=0.73$. This implies a scale of 7.84 kpc~arcsec$^{-1}$ at the redshift of El Gordo and a luminosity distance of 5568.7 Mpc \citep{Wright06}.

\begin{table*}
	\centering
	\caption{Summary of uGMRT observations. }
	\label{tab:obstable}
	\begin{tabular}{lcccccc} 
		\hline
		Frequency  & Band Width&Date & Duration & robust & Beam&rms\\
        Band   & (MHz)&     & hours & &$\prime\prime \times \prime\prime$, position angle&$\mu$Jy~beam$^{-1}$\\
		\hline
       	100 - 500 MHz& 400$^\dag$	& 16 Nov 2017  & 5& 0&$14.5''\times5.9''$, $3.5^{\circ}$ &50\\
       	       	\hline
		550 - 950 MHz & 400 & 20 Nov. 2017 & 5 &0&$9.6''\times3.8''$, $7.6^{\circ}$ &13 \\
\hline
		1050 - 1450 MHz& 400 & 09, 10 Nov. 2017 & $5+5$ &0&$6.2''\times2.4''$, $-$0.82$^{\circ}$ & 15\\
		\hline
	\end{tabular}
 \tablefoot{$^\dag$ The usable band was within 300 - 500 MHz.}
\end{table*}


\section{uGMRT observations and data analysis}\label{ugobs}

El Gordo was observed with the uGMRT under the proposal code $33\_030$ in 
bands 3 (300 - 500 MHz), 4 (550 - 950 MHz), and 5 (1050 - 1450 MHz). A summary of the observations used for analysis is provided in Table~\ref{tab:obstable}. At band-5 we used observations from two observing sessions. At band-4, the GMRT antennas C00, C14, S04, and E04 were not equipped with the wide-band uGMRT feeds at the time of observations and thus provided data only in a narrow band of 100 MHz. These antennas were removed from the data at band-4. 

We used the data analysis pipeline for the uGMRT called {\bf{C}}ASA {\bf{P}}ipeline-cum-{\bf{T}}oolkit for {\bf{u}}GMRT Data {\bf{Re}}duction (\texttt{CAPTURE\footnote{\url{https://github.com/ruta-k/CAPTURE-CASA6}}}, \citealt{2021ExA....51...95K}). The standard steps of flagging the data, flux, and bandpass calibration using the primary calibrator, complex gain calibration using the secondary calibrator, followed by application of the calibrator solution to the target were performed using the pipeline. 
We have used the \texttt{Perley-Butler 2017} flux density scale \citep{2017ApJS..230....7P} for absolute flux density calibration. The data after calibration were further examined for the presence of bad data and consequently flagged. After this flagging, the data were re-calibrated and solutions were applied on the target source. The target source data were then split into a separate file and further flagging was carried out. These data were then averaged in frequency channels to reduce the volume of the data while still not being affected by bandwidth smearing. For the imaging and self-calibration step the released \texttt{CAPTURE} pipeline was modified as follows. The frequency-averaged target source visibilities were split into 4 - 6 sub-bands and a combined image was produced. The gain calibration was carried out separately for the sub-bands but the imaging was done for the combined file using the \texttt{CASA} task \texttt{tclean} with the choice of \texttt{robust$ = 0$} and the multi-term (\texttt{nterms$ = 2$}) multi-frequency wide-field method. Typically 4 - 6 rounds of phase self-calibration followed by 3 rounds of amplitude and phase self-calibration were carried out. For band-5, the self-calibrated visibilities were combined from the two observing sessions and a joint image from the data was made using \texttt{tclean}.

We used \texttt{CASA} for obtaining the flux densities of radio sources in the fields. The error on absolute flux calibration, $\sigma_{\rm abs}$ of $10\%$ is used for all the bands \citep{2017ApJ...846..111C}. The error on the flux density of discrete sources was obtained by adding the fitting error in quadrature with the $\sigma_{\rm abs}$. The error on the flux density of extended sources was calculated according to $\sqrt{(\sigma \sqrt N_{\mathrm b})^2 + (\sigma_{\mathrm{abs}} S_{\nu})^2}$,
where $N_{\mathrm b}$ is the number of beams in the extent of the emission and $\sigma$ is the rms noise. All the images were corrected for the primary beam attenuation.

\section{X-ray Data analysis} 
\label{xrayobs}

We made use of the same \textit{Chandra} data reduced and originally presented in \citet{2016MNRAS.463.1534B} to produce thermodynamical maps of the ICM of El Gordo. The data consists of three \texttt{ACIS-I VFAINT} observations (ObsID: 12258, 14022, 14023) accounting for a total net exposure time of 340 ks. In order to make the thermodynamical maps, we created spectral extracting regions deploying the \texttt{CONTBIN} binning algorithm \citep{2006MNRAS.371..829S}, which aims to generate regions following the surface brightness of the X-ray emission. For each region, spectra were thus extracted from all ObsIDs and fitted simultaneously in \texttt{XSPEC} \citep{1996ASPC..101...17A} adopting the background model used in \citet{2016MNRAS.463.1534B} and a thermal APEC model, with a fixed metallicity of 0.3 solar, for the ICM emission. Temperature is a direct result of the fitting, while values for pseudo-pressure and pseudo-entropy were calculated as $kT \times A^{1/2}$ and $kT \times A^{-1/3}$, respectively, where $A$ is the APEC normalization \citep[see e.g.][]{2007A&A...463..839R, 2012MNRAS.423..236R, 2018MNRAS.476.5591B}.

\begin{figure}
\centering
  \includegraphics[height = 6.5 cm]{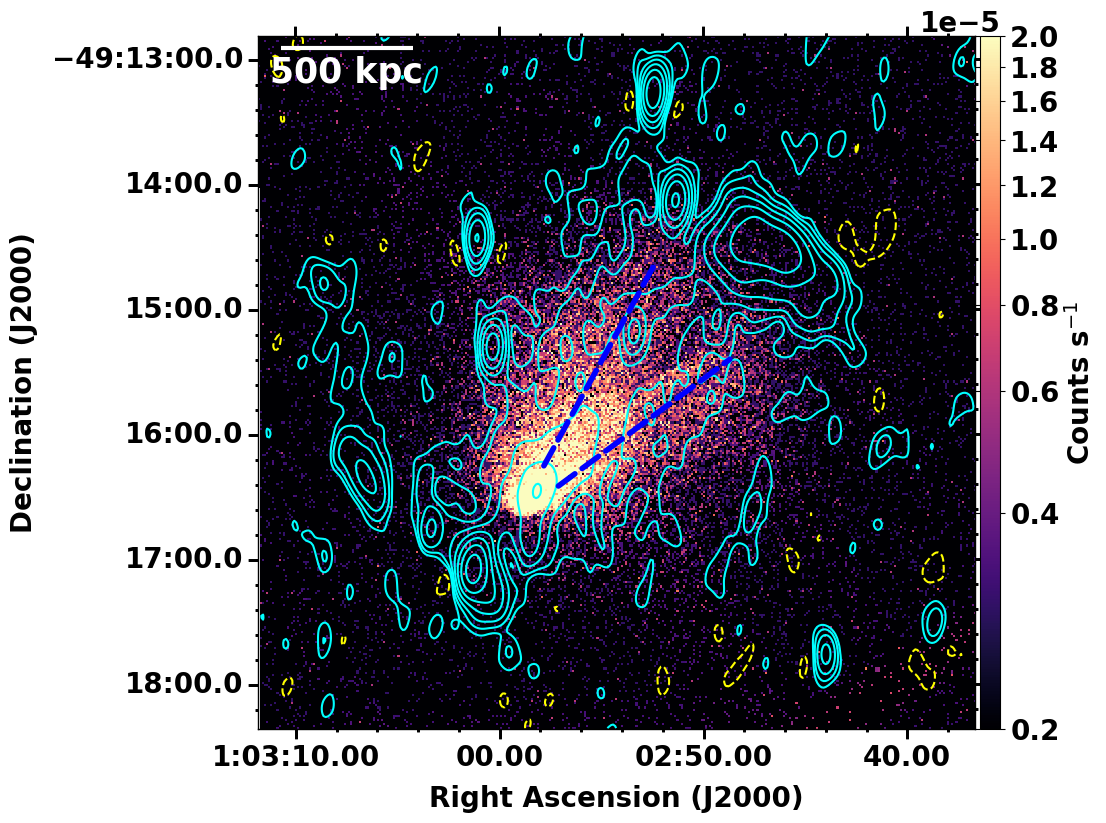}
    \caption{uGMRT band-3 image is shown in contours overlaid on the point source-subtracted \textit{Chandra} X-ray image in colour. The contour levels are $-$0.1, 0.1, 0.2, 0.4,... mJy~beam$^{-1}$. The positive radio contours are shown by the solid line (cyan) and the negative as dashed line (yellow). The synthesized beam size of the band-3 image is $14.5''\times5.9''$, position angle $3.5^{\circ}$. The blue dashed segments indicate the northern and southern X-ray tails, respectively.}
    \label{eg-b3-fig1}
\end{figure}

\begin{figure}
    \centering
        \includegraphics[height=9.8cm]{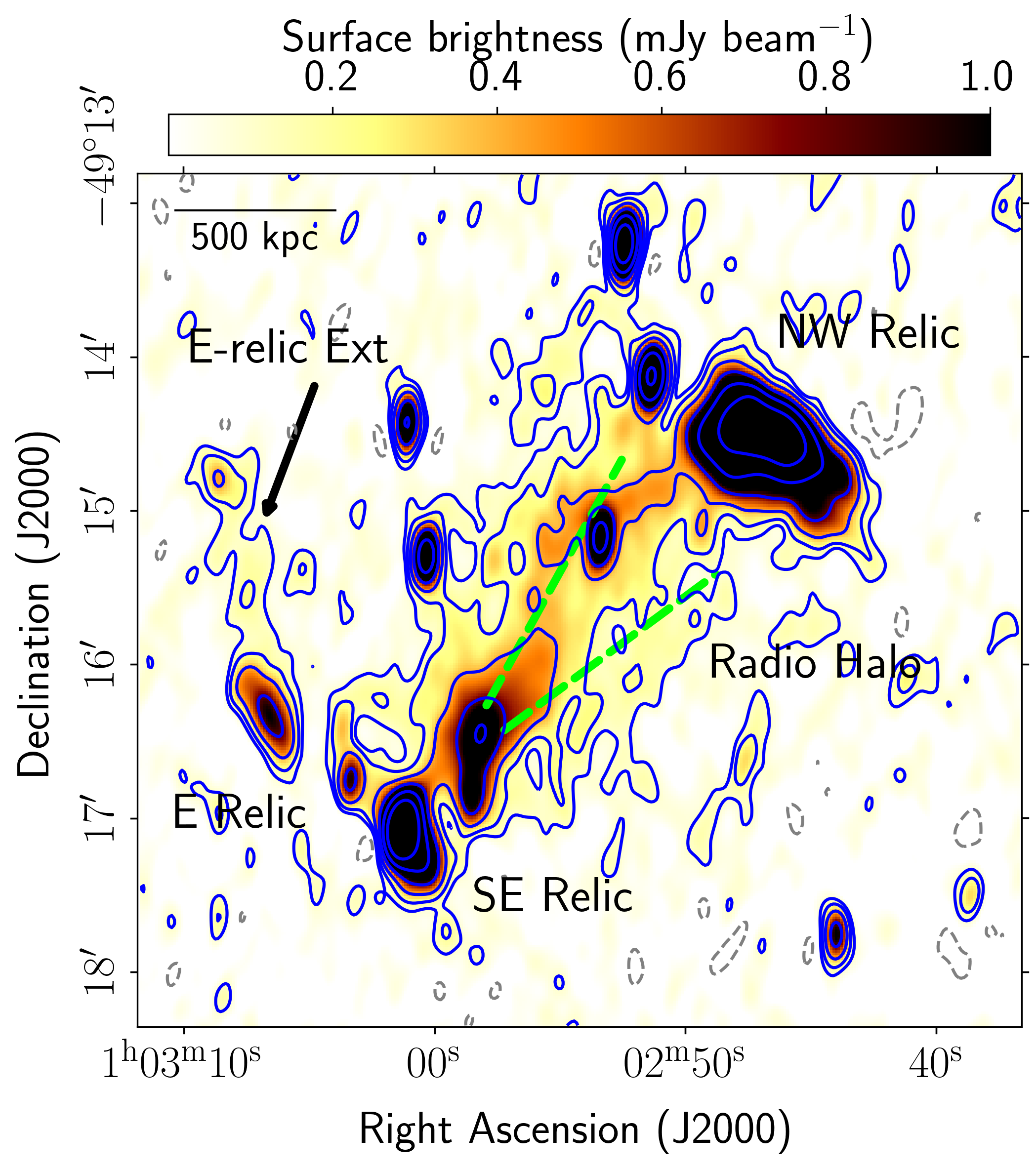}
    \caption{uGMRT band-3 image is shown in color and in contours. The contour levels are $-$0.1, 0.1, 0.2, 0.4,... mJy beam$^{-1}$. Blue solid lines are positive contours and negative contour is shown in grey dashed line. The beam size is $14.5''\times5.9''$, position angle $3.5^{\circ}$. The diffuse sources are labeled. The arrow points out the new further extended feature of the E-Relic. The dashed segments mark the two X-ray tails.}
    \label{eg-b3-srcs}
\end{figure}

\section{Results}\label{results}

\subsection{Radio images}\label{images}

\begin{figure*}[ht]
    \centering
    \begin{minipage}{0.49\textwidth}
        \centering
       \includegraphics[height=10cm]{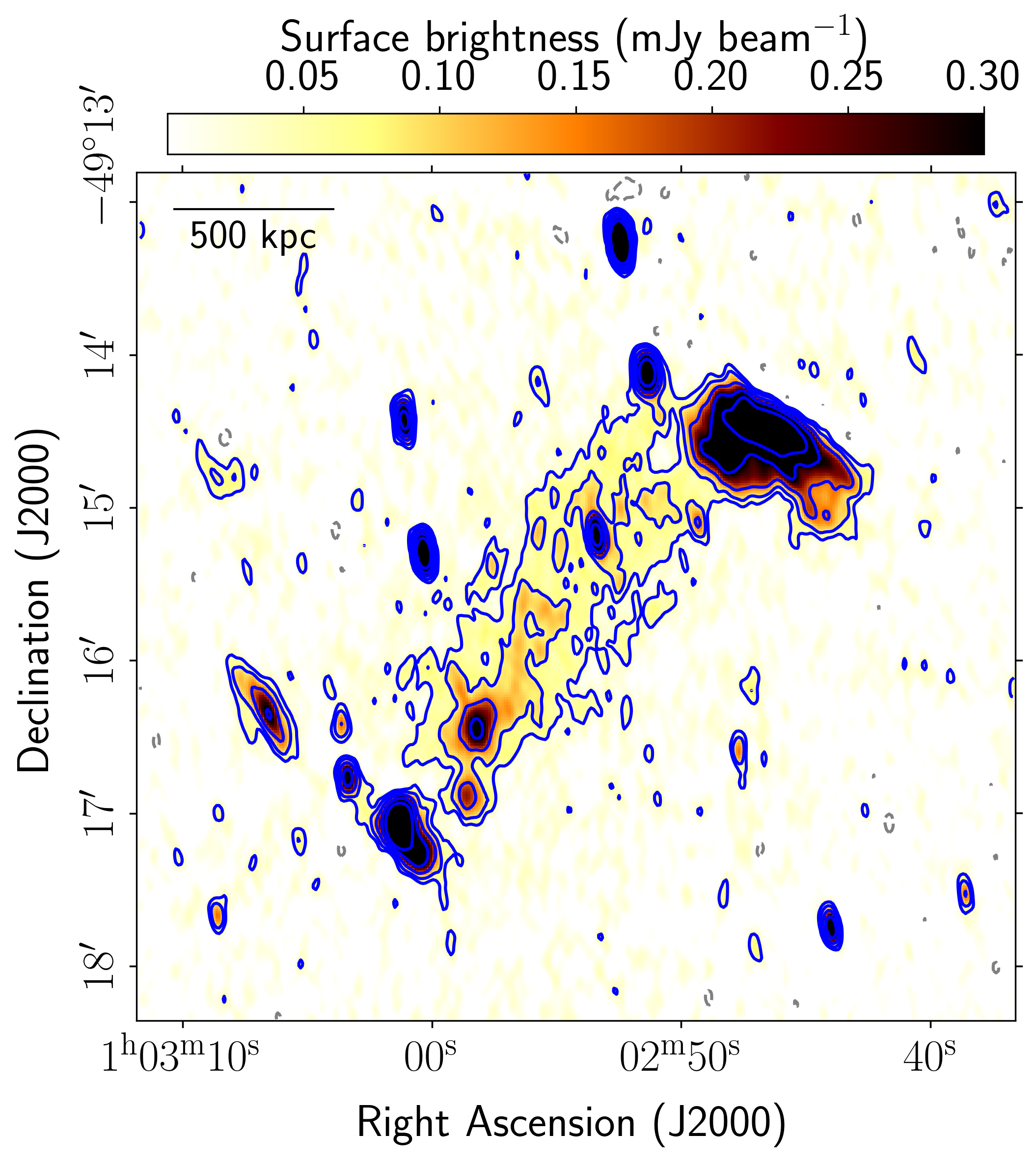}
    \end{minipage}
    \hfill
    \begin{minipage}{0.49\textwidth}
        \centering
       \includegraphics[height=10cm]{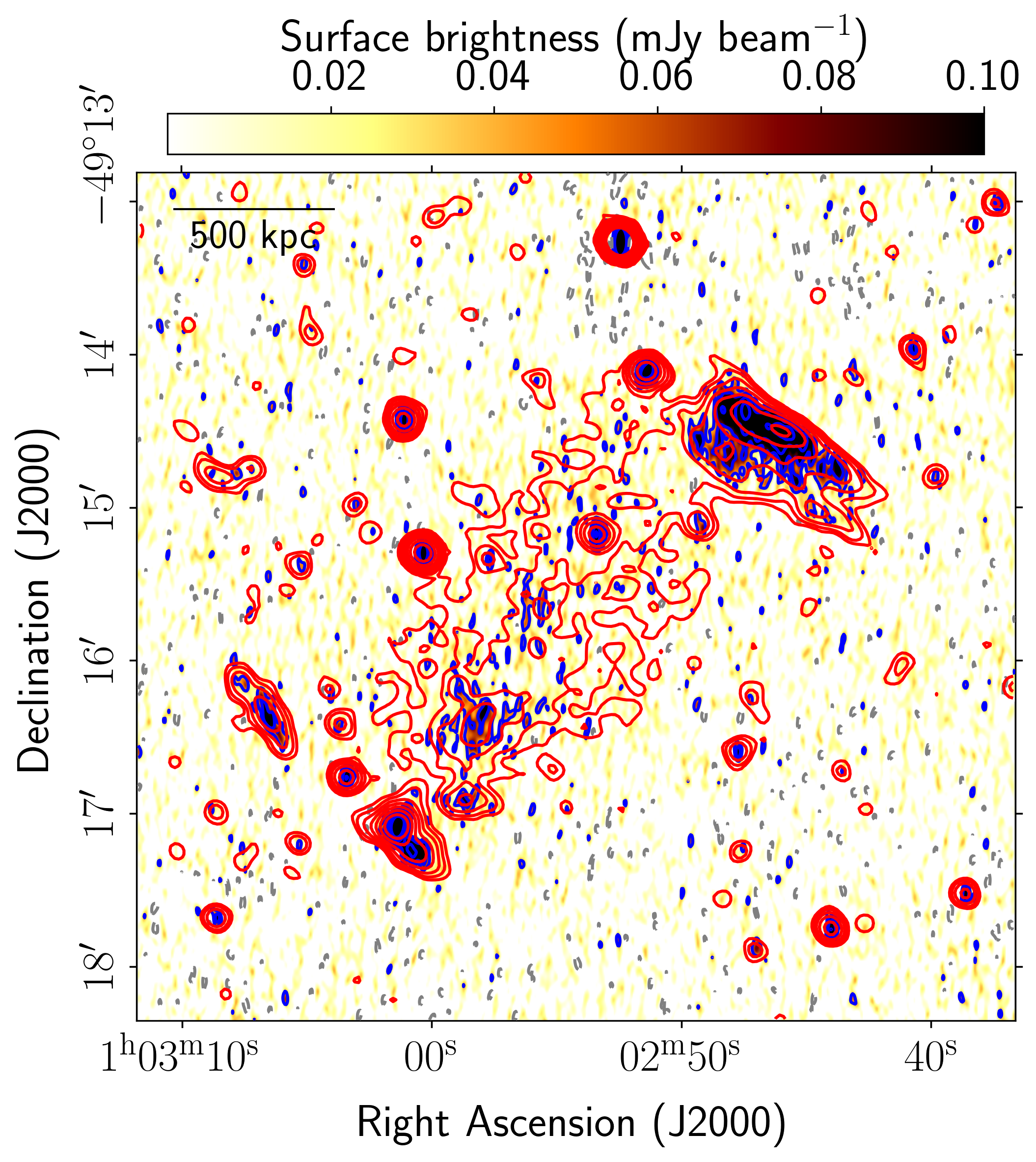}
    \end{minipage}
    
    \caption{\textit{Left:} uGMRT band-4 image shown in color and in contours. The beam is $9.6''\times3.8''$, position angle $7.5^{\circ}$. \textit{Right:} uGMRT band-5 image is shown in color and in contours (blue positive and grey negative). The beam is $6.1''\times2.4''$, position angle of $-0.81^{\circ}$. The contour levels in both panels are $-$0.04, 0.04, 0.08, 0.16, ... mJy beam$^{-1}$. The MeerKAT 1.28 GHz image \citep{knowles22} with a resolution of $7.2''\times6.6''$, position angle $44.6^\circ$ is overlaid in red contours at the levels of $15, 30, 60, ...$ $\mu$Jy beam$^{-1}$. The rms in the MeerKAT image is $5 \mu$Jy beam$^{-1}$.}
    
    \label{eg-b4b5}
\end{figure*}

\begin{figure}
    \centering
  \includegraphics[height=9.8cm]{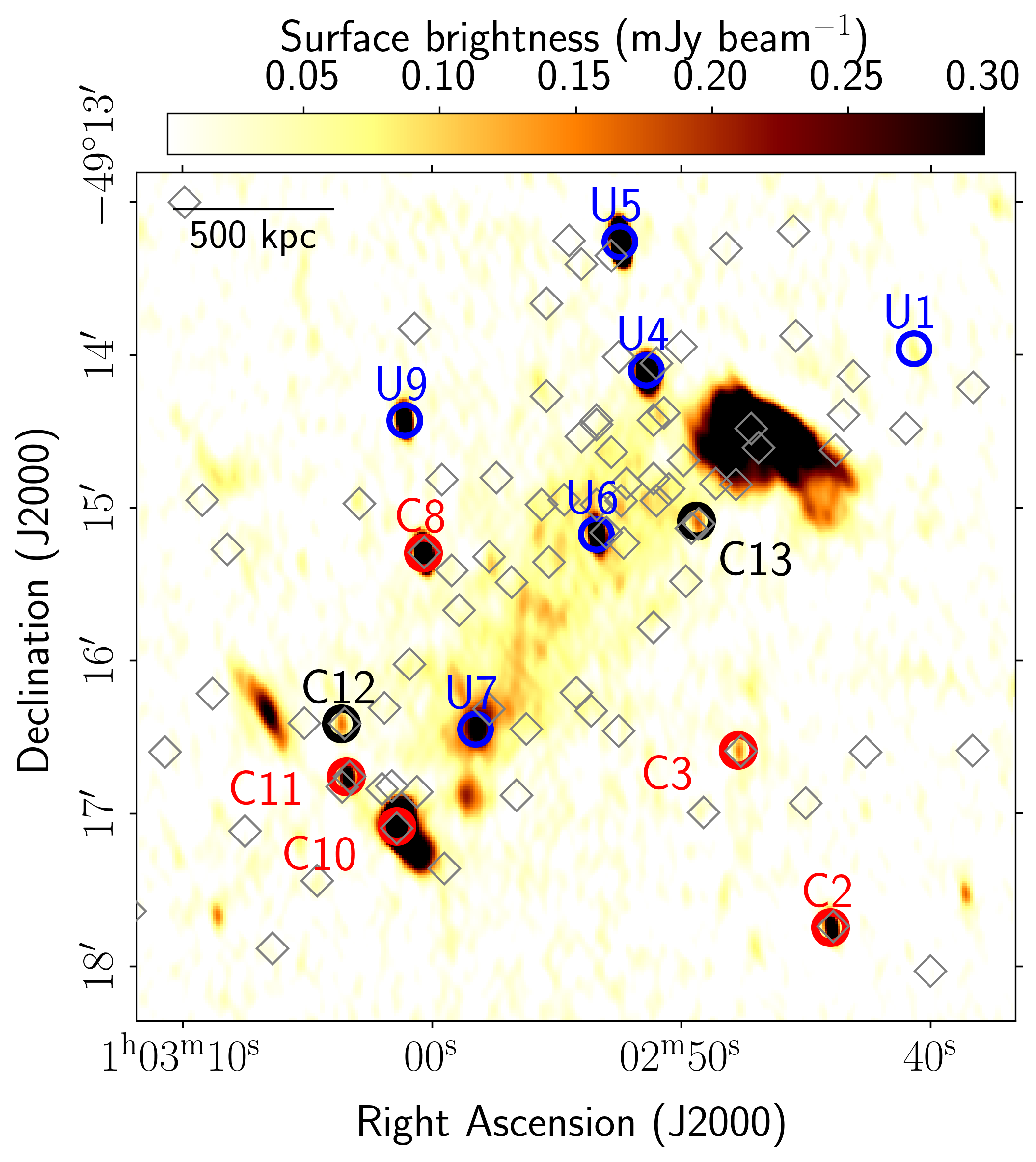}
        \caption{uGMRT band-4 image in color is shown with the discrete sources marked as circles and labeled. The diamonds show the positions of the spectroscopically identified cluster members by \citet{sifon2013}. The source labels from L14 are shown and two new labels C12 and C13 are added. The prefix `U' refers to sources unrelated to the cluster and those with `C' are cluster members.}
    
    \label{eg-ptsrc}
\end{figure}

We present the image of El Gordo using uGMRT band-3 overlaid on the \textit{Chandra} X-ray image in Figure~\ref{eg-b3-fig1}. The radio emission is co-spatial with the X-ray emission and is elongated in the same direction from northwest to southeast. In Figure~\ref{eg-b3-srcs}, we show the band-3 image with labels for the diffuse emission from L14. The additional radio emission detected for the first time near the E-Relic is labeled as ``E-relic Ext''. The uGMRT band-4 and band-5 images are presented in Figure.~\ref{eg-b4b5}, where the MeerKAT image at 1.28 GHz \citep{knowles22} is overlaid on the uGMRT band-5 emission for comparison. There is a good match between the uGMRT band-5 sources and the MeerKAT image. The discrete sources in the field with labels from L14 are shown in Fig.~\ref{eg-ptsrc}, on the uGMRT band-4 image shown in colour. The sources with the prefix ``C'' belong to El Gordo and those with the prefix ``U'' are unrelated. The sources labeled C12 and C13 are discrete sources associated with the spectroscopically identified galaxies \citep{sifon2013} in El Gordo that we found in addition to those listed by L14. The measured flux densities of the discrete sources at the three bands are given in Table ~\ref{radsrc}.

\begin{table*}
	\centering
	\caption{Properties of the discrete radio sources in the cluster field. The labels prefixed `C' are cluster members identified by \citet{sifon2013} and with `U' are unrelated to the cluster. 
	\label{radsrc}}
	\begin{tabular}{ccccccl} 
		\hline
Cluster/Source	& RA$_{\rm J2000}$	& Dec$_{\rm J2000}$& $S_{1274 \rm{MHz}}$	&$S_{672\rm{MHz}}$ & $S_{366\rm{MHz}}$ \\
Label        &  hh mm ss     & $^{\circ}$ $'$ $''$ &   mJy & mJy &mJy                \\
\hline
U1&01 02 40.594 & $-$49 13 59.97 & $0.20\pm0.04$ &-&-\\
C2 &01 02 43.975 & $-$49 17 45.48 & $0.43\pm0.04$ &$0.68\pm0.07$&$1.22\pm0.14$\\
C3&01 02 47.725 & $-$49 16 35.90 &$0.11\pm0.02$ &$0.17\pm0.03$&$0.85\pm0.17$ \\
U4&01 02 51.361 & $-$49 14 06.55 &$1.28\pm0.09$ &$2.21\pm0.25$ &$4.27\pm0.86$\\
U5&01 02 52.435 & $-$49 13 15.97 &$5.69\pm0.29$ &$5.93\pm0.59$&$6.48\pm0.66$\\
U6&01 02 53.393 & $-$49 15 10.69 &$0.77\pm0.05$ &$1.31\pm0.13$&$2.05\pm0.24$\\
U7&01 02 58.136 & $-$49 16 24.68 &$0.65\pm0.15$ &$1.24\pm0.17$&$3.40\pm0.50$\\
C8&01 03 00.320 & $-$49 15 17.98 &$1.94\pm0.10$ &$2.12\pm0.21$&$2.65\pm0.29$\\
U9&01 03 01.090 & $-$49 14 25.82 &$0.86\pm0.06$ &$1.12\pm0.11$&$2.03\pm0.22$\\
C10&01 03 01.365 & $-$49 17 05.19 &$3.37\pm0.21$ &$6.73\pm0.74$&$15.50\pm1.80$\\
C11&01 03 03.393 & $-$49 16 45.86 &$0.43\pm0.04$ &$0.60\pm0.15$&$3.00\pm1.80$\\
C12&01 03 03.470 & $-$49 16 26.19 &$0.11\pm0.05$ &$1.24\pm0.17$&$0.82\pm0.16$\\
C13&01 02 49.249 & $-$49 15 08.17 &$0.17\pm0.04$ &$0.18\pm0.05$&$0.65\pm0.15$\\
\hline
\end{tabular}
\end{table*}

\begin{table*}
	\centering
	\caption{Flux densities of the diffuse sources in El Gordo. The spectral indices ($\alpha$) are from the fits shown in Fig.~\ref{intspec}. The largest linear sizes (LLS) are measured from the band-3 image presented in Fig.~\ref{eg-b3-srcs}.
	}
	\label{tab:intspec}
	\begin{tabular}{lccccccc} 
		\hline
		Radio    & S$_{366\mathrm{MHz}}$    & S$_{672\mathrm{MHz}}$  & S$_{1274\mathrm{MHz}}$  & $\alpha$  & $P_{1.4 \rm GHz}$ &LLS\\
        Emission &mJy & mJy & mJy & & $10^{25}$ W Hz$^{-1}$&kpc\\
		\hline
         Radio Halo &$34.3\pm3.5$ &$16.0\pm1.6$ &$8.7\pm1.0$ & $-1.0\pm0.3$&$3.20\pm0.20$&1230\\
         NW Relic  &$51.2\pm5.1$  &$23.0\pm2.3$  &$10.5\pm1.0$& $-1.3\pm0.4$&$4.50\pm0.20$&840\\
         SE Relic &$4.0\pm0.4$ &$2.5\pm0.3$  & $1.10\pm0.17$& $-1.4\pm0.4$&$0.47\pm0.07$&800\\
         E Relic & $3.9\pm0.4$ & $1.9\pm0.2$ &$0.9\pm0.2$& $-1.4\pm0.2$&$0.44\pm0.08$ &430\\
         E Relic Ext &  $2.4\pm0.2$        &      $0.7\pm0.1$      &  $<0.15$  & $-2.1\pm0.4$&$0.38\pm0.07$&600 \\    
		\hline
	\end{tabular}
\end{table*}

\subsection{Integrated spectra of the diffuse components}\label{intspectra}

\begin{figure}
    \centering
            \includegraphics[trim =0.0cm 0.5cm 1cm 1cm,clip,height = 8.5 cm]{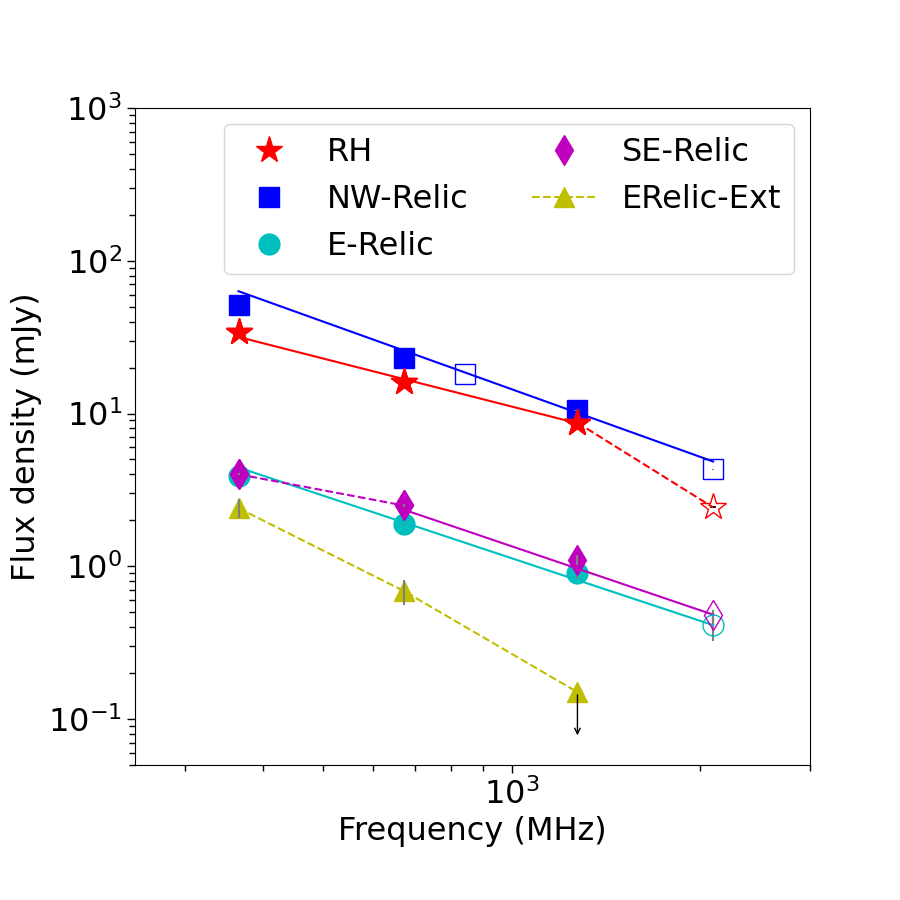}
            \caption{Integrated spectra of the diffuse sources in El Gordo. The open symbols show the points from L14.
            The solid lines show the fitted power laws. The spectral indices are reported in Table~\ref{tab:intspec}. The dashed lines join the points for the respective diffuse components that were not included in the fits.}
    \label{intspec}
\end{figure}

The radio relics and the halo likely have structures on angular scales smaller than the resolution. Comparison of the images at bands 3, 4 and 5 reveal that there are structures within the relic and radio halo even at the highest resolution images at band 5. At the redshift of El Gordo, the beam of 10$''$ corresponds to 78 kpc which is in the range where relics and the halo can have emission. Thus subtracting point sources using uv-range cutoff would subtract the features in the halo and the relics. The flux densities of diffuse emission were measured from the \texttt{robust 0} images convolved to a common resolution of 16$''$ resolutions in matched regions  (Fig.~\ref{flux-regions}). The flux densities of the discrete sources measured in robust 0 images made with uv-range cutoff ($>3.4 k\lambda$) were then subtracted from these. 
The E-Relic flux density for the part already reported in L14 and the extension we see in the bands 3 and 4 images are given separately. The flux densities and the spectral indices obtained from power law fits are reported in Table ~\ref{tab:intspec}. The emission from the SE relic is well-fitted with a single power law, however, marginal steepening is seen. The integrated spectra and the power law fits are shown in Figure~\ref{intspec}. The k-corrected radio power of the radio halo at 1.4 GHz is $(3.20\pm0.20)\times10^{25}$ W~Hz$^{-1}$. The largest linear sizes of the diffuse sources measured at band-3 are reported in Table~\ref{tab:intspec}. We compared the band 5 flux densities (discrete sources and diffuse emission) with those in the MeerKAT 1.28 GHz image and found them to be consistent within $1\sigma$ uncertainties. A study with matched uv-coverage and removal of discrete sources at uGMRT bands 3 and 4 and MeerKAT 1.28 GHz will be undertaken in future.

\subsection{Spectral index map}\label{specmap}

\begin{figure*}
    \centering
            \includegraphics[height=7.0cm]{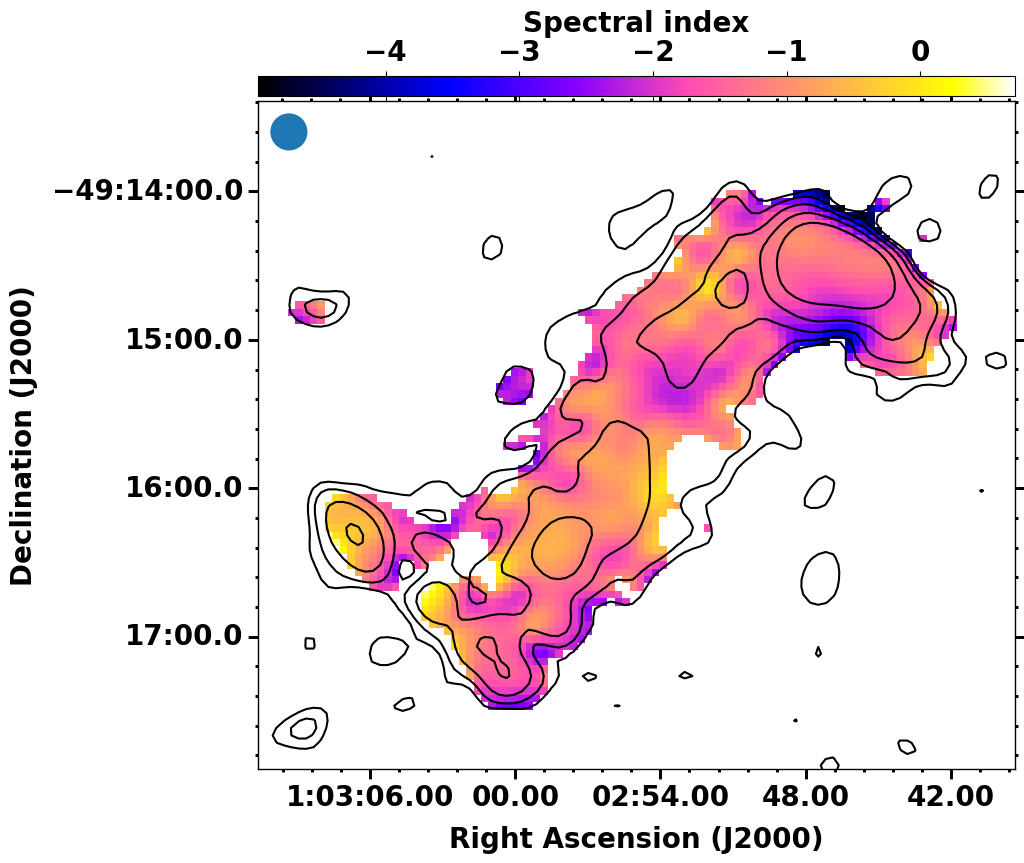}
           \includegraphics[height = 7.0 cm]{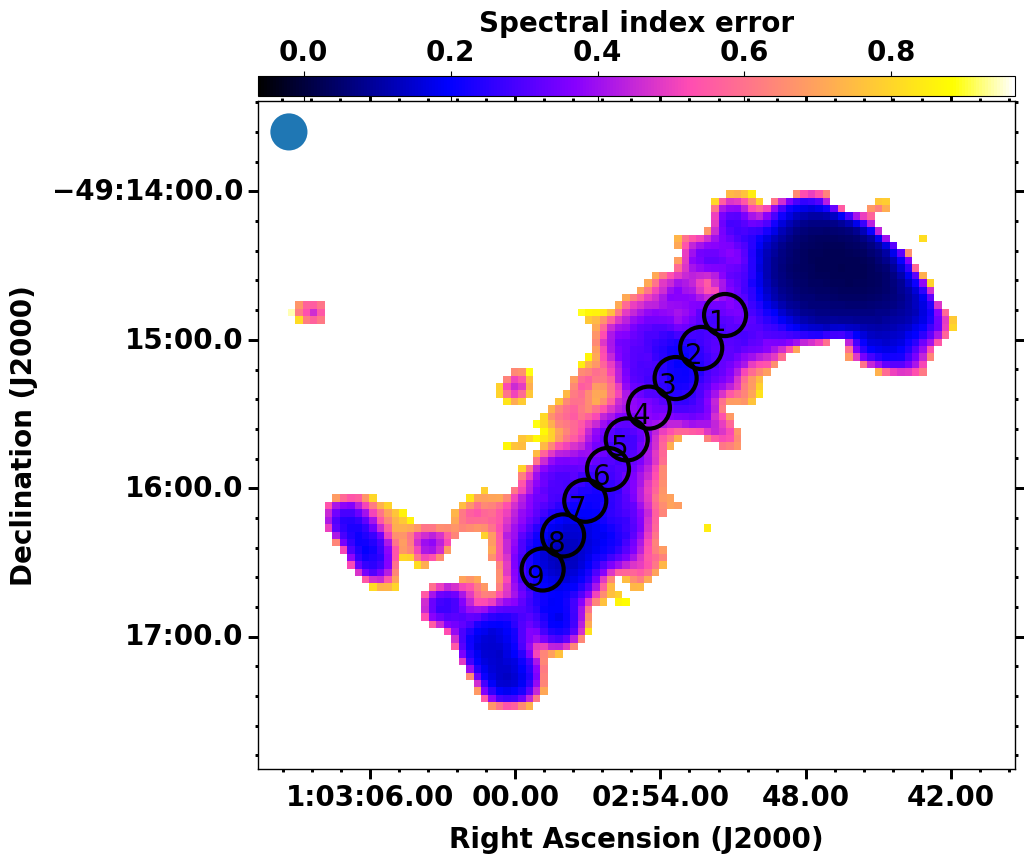}   
        \caption{\textit{Left:} Spectral index map  between 366 MHz and 672 MHz for the radio halo in El-Gordo at 15$'' \times 15''$ resolution is shown in colour scale. The contours from 672 MHz image at 0.1, 0.2, 0.4, 0.8 and 1.6 mJy beam$^{-1}$ are overlaid. \textit{Right:} Spectral index error map is shown. High errors at the edges are expected due to the low S/N at the outskirts. The numbered regions used for analysis are shown on the map. The beam is shown at the top left in both the panels.}
    \label{spix}
\end{figure*}

We have created a spectral index map using band-3 and 4 observations where the maximum extent of the radio halo was detected at both frequencies. We used a \textit{uv}-range cut of \textgreater 3.4k$\lambda$ to image the contribution from the discrete sources. Here the mask was manually created to avoid the part of the NW-relic where there is structure even at the scales of discrete sources. The discrete source model was subtracted from the \textit{uv}-data and the residual visibilities were imaged using uniform weights in the \textit{uv}-range of $0.065-12$k$\lambda$ at both the bands. The images were convolved to the highest possible common circular resolution of $15'' \times 15''$. These were then combined to make the spectral index map and the corresponding error map shown in Figure.~\ref{spix}. We have masked the pixels below 3$\sigma_{\rm rms}$ from each image. This helps to avoid regions with large errors on the spectral index.

\begin{figure}
    \centering
        \includegraphics[width=8.5cm]{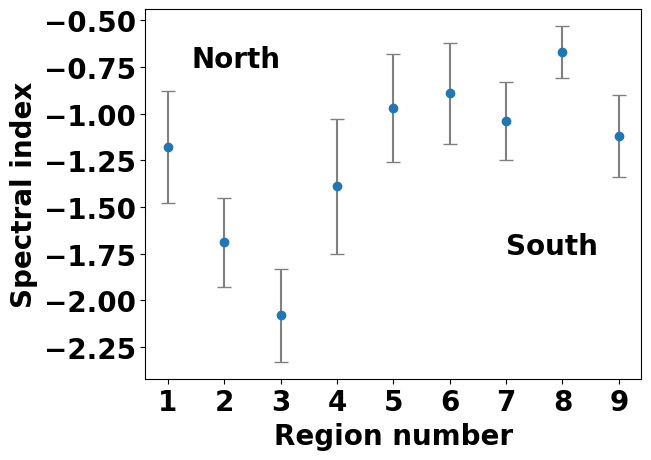}
            \caption{The spectral indices with error bars obtained from the spectral index map are plotted as a function of the numbered regions shown in Fig.~\ref{spix}, right.}
    \label{halo-spix-plot}
\end{figure}

The spectral index trend from north to south within the radio halo was analyzed in the numbered regions marked in Fig.~\ref{spix}, right. In fig.~\ref{halo-spix-plot}, the trend of the spectral index over the regions is shown. While in the northern part the spectral indices reach values $<-1.5$, the southern region is flatter than $-1.2$.
In the NW-relic there is an overall steep to flat spectral index trend from the outer to the inner edge. 
There is marginal evidence of steepening of spectral indices in the SE-relic. Near the E-relic, the spectra show a flatter outer edge and there is marginal evidence for steepening. The resolution of the map is not sufficient to resolve the spectral index across the E-relic as well as for the extension.

\subsection{Radio and X-ray point to point analysis}\label{sec:ptp}

\begin{figure*}
    \centering
    \includegraphics[height=6cm]{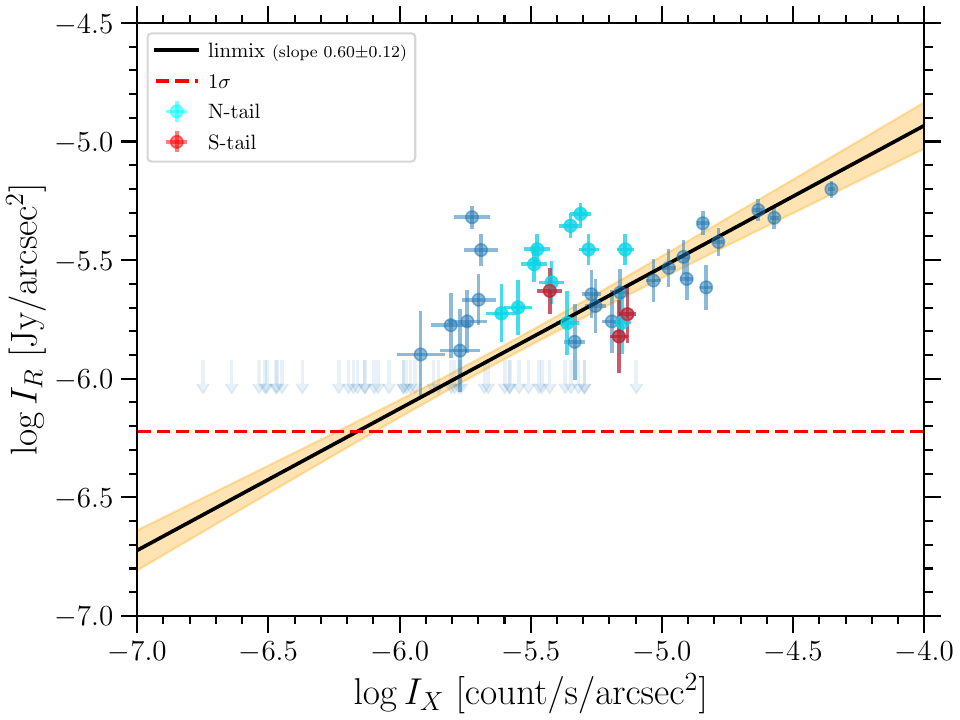}
    \includegraphics[height=6cm]{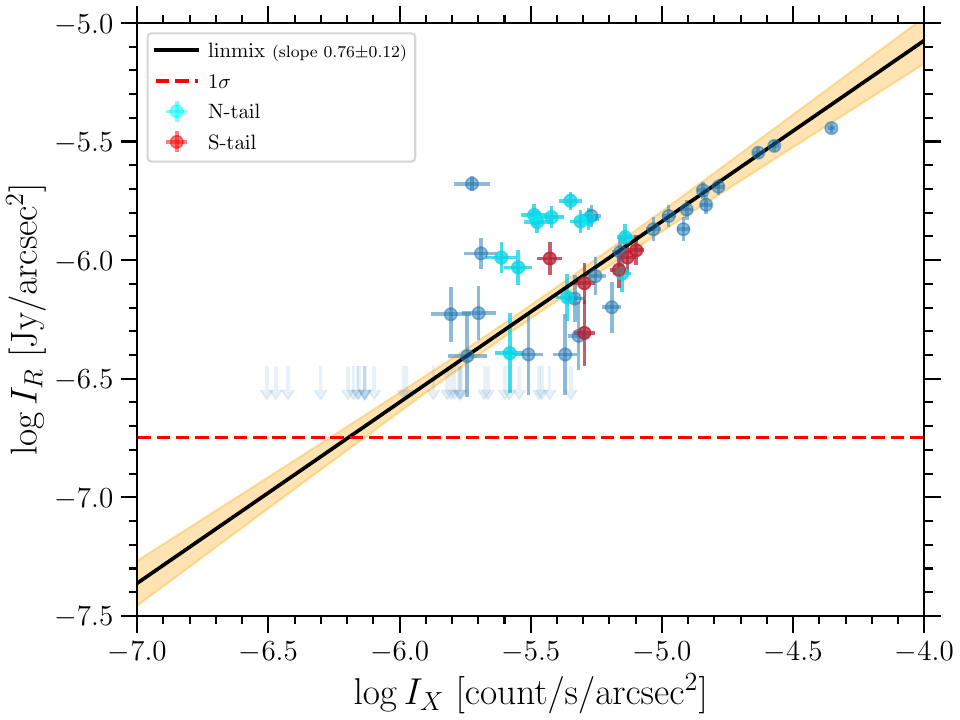}
    \caption{\textit{Left:} The radio surface brightness ($I_R$) versus the X-ray surface brightness ($I_X$) is plotted for the band 3. The points in these plots are from the grid shown in Figure~\ref{ptpgrid}. The cyan points are from the northern X-ray tail, red points from the southern X-ray tail and the remaining from the rest of the halo. The fit to the points is shown in a black solid line with the 95\% confidence region shown in the shaded part. The red dashed line shows the $1\sigma$ level of the radio images. The light blue arrows indicate the upper limits, above 2$\sigma_{\rm rms}$ points. \textit{Right:} The $I_R$ versus $I_X$ scaling plot for band 4 in the same colour code as the left panel.}
    \label{ptp}
\end{figure*}

The morphology of radio halos generally follows that of the X-ray emission, suggesting an interplay between thermal and non-thermal components in the ICM \citep[e.g.,][]{2001A&A...369..441G, giacintucci05,cova19,xie20,rajp21,balboni24}. This can be observed also in El Gordo, where both the radio and X-ray emissions are elongated along the NW-SE direction, (together with the relics) indicating the main merger axis (Figure.~\ref{eg-b3-fig1}). To investigate the connection between thermal and non-thermal components more quantitatively, we used the \textit{Chandra} and uGMRT data to compare the X-ray and radio surface brightnesses on a point-to-point basis \citep[e.g.,][]{2001A&A...369..441G}. 

As the first step of the analysis, we constructed a grid that covers the cluster where each element of the grid is constituted by a square cell with a width of 9 pixels (1 pixel = 2$''$), broader than the synthesized beam (15$''$) of the radio image. We then measured the X-ray surface brightness in each cell from the \textit{Chandra} point-source subtracted image and the radio surface brightness from the band 3 and band 4 images used to produce the spectral index map. As we are interested in the radio halo emission, we excluded the cells located in the directions of the two relics from the analysis. The grid obtained in this way is shown in Figure~\ref{ptpgrid} (in Appendix~\ref{grid}). 

In the second step, we plot the radio and X-ray surface brightnesses in the $I_R-I_X$ plots (Figure~\ref{ptp}) to search for possible correlations between the two quantities. Generally, a power law relationship in the form

\begin{equation}
    \log I_R = b \log I_X + A
\end{equation}

is used to fit the data, where the slope of the scaling ($b$) determines whether the radio brightness decreases more slowly than the X-ray brightness (if $b<1$) or vice versa (if $b>1$). We restricted the analysis and fitting to the data points where the radio surface brightness is $>3\sigma$, and considered as upper limits (cyan arrows in Figure.~\ref{ptp}) the values below this threshold, following \citet{2020ApJ...897...93B} \citep[see also][]{2021ApJ...907...32B,2021arXiv210405690R}.
This cut on the $y$-axis requires the usage of a Bayesian linear regression method to take into account the selection effect. Therefore, we adopted the \texttt{linmix} package \citep{2007ApJ...665.1489K}, which can handle censored data, to perform linear regression. The best-fit slopes for the band 3 and band 4 data are $b=0.60\pm0.12$ and $b=0.76\pm0.12$, respectively, suggesting that the magnetic field strength and relativistic particle density (traced by $I_R$) declines slower than the thermal gas density (trace by $I_X$). The obtained slopes are in line, with what is seen for the giant radio halos \citep[e.g.,][]{rajp23,2024A&A...686A...5B,santra24b}. The hint of an increase in $b$ with frequency could be due to spectral steepening in fainter portions (outskirts) in X-rays. The strength and monotonic behavior of the relationship is measured using the Pearson and Spearman correlation coefficients for the band 3 and band 4 data to be $r_p=0.54$ and $r_s=0.47$, and $r_p=0.64$ and $r_s=0.56$, respectively.

\begin{figure}
    \centering    
    \includegraphics[width=\columnwidth]{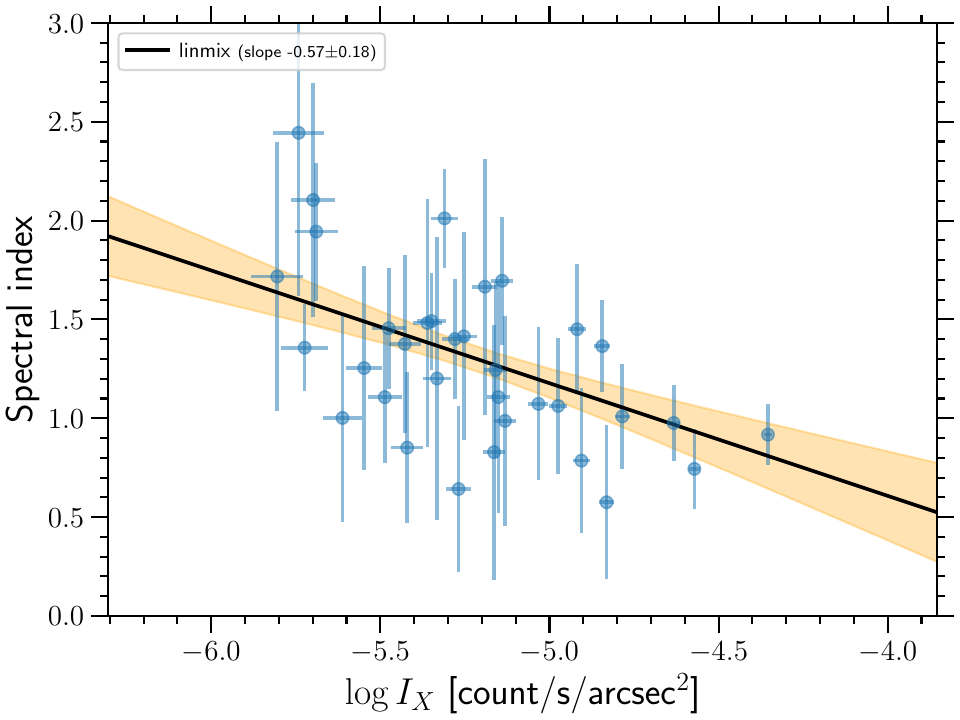}
    \caption{The spectral index ($\alpha$)-X-ray surface brightness point-to-point comparison is shown in blue filled circles. The error bars are statistical measurement errors. The best fit is shown with the black solid line with the confidence interval shown as the shaded region.}
    \label{ptp-alpha}
\end{figure}

A comparison of the spectral index distribution over the radio halo and the distribution of X-ray surface brightness can be investigated to check for any correlation. For this purpose, a point-to-point comparison of the spectral index and X-ray surface brightness was carried out. The same grid as shown in Figure~\ref{ptp-alpha} was used but only the cells where the surface brightness in both, bands 3 and 4, exceeded 2-$\sigma$ were used (34 cells). The scatter found in the spectral index ($\alpha$) and X-ray surface brightness ($I_X$) was investigated for the presence of a correlation. The Pearson's and Spearman's correlation coefficients of $-0.58$ and $-0.57$, respectively, were found. With the \texttt{linmix} package the slope of the fitted line was found to be $-0.57\pm0.18$. The spectral index and X-ray surface brightness show an anti-correlation, implying the fainter parts in X-rays have steeper spectral indices, and a radially outward spectral steepening. 

\subsection{Comparison with X-ray temperature, pressure and entropy}\label{sec:xraytemp}
\begin{figure*}
    \centering
\includegraphics[height = 8.5cm]{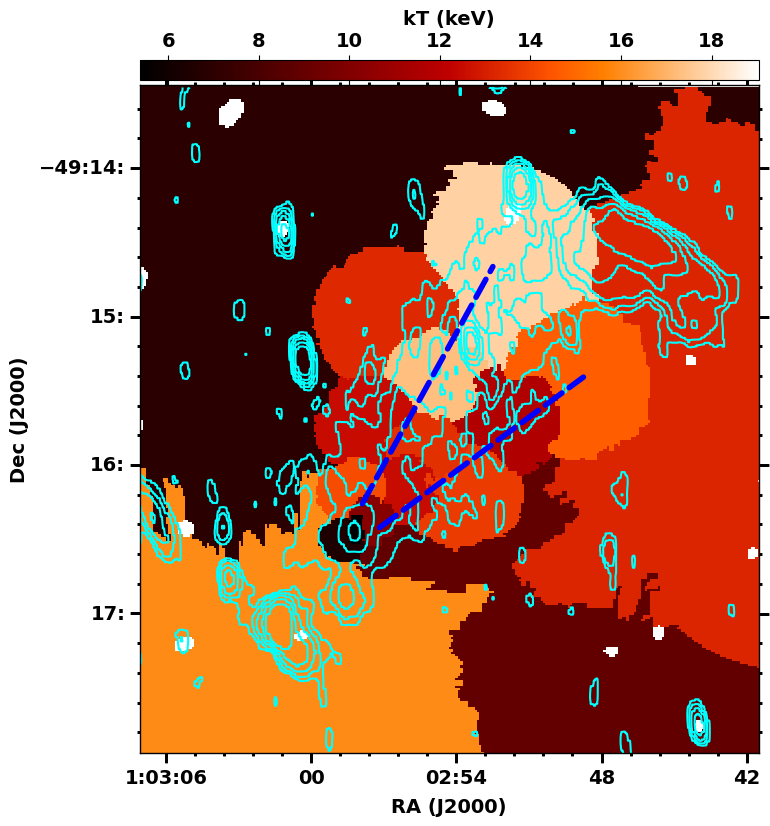}
\includegraphics[height=8.5cm]{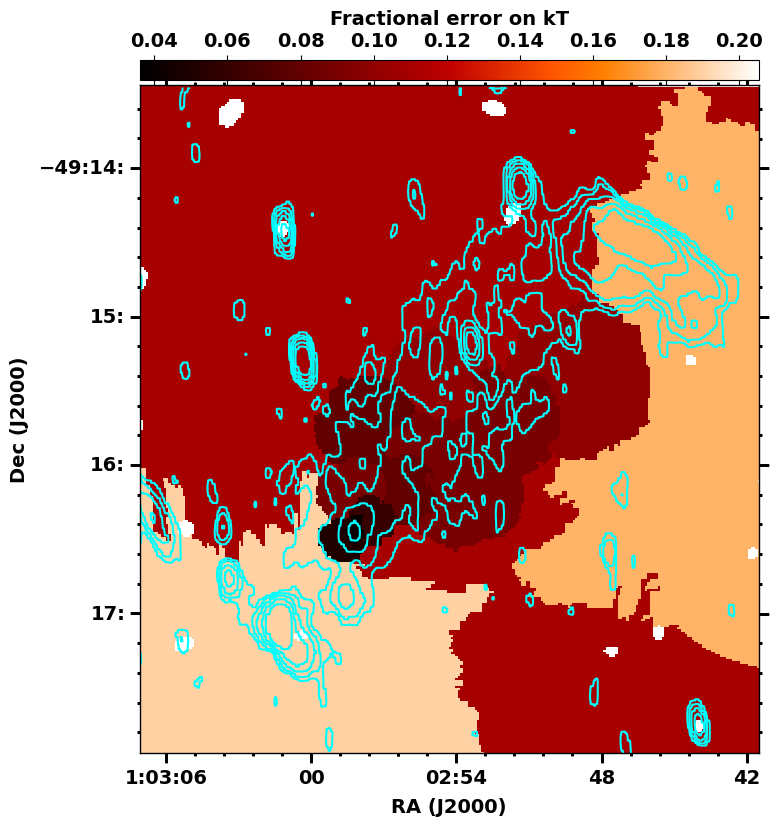}
\includegraphics[height = 8.5cm]{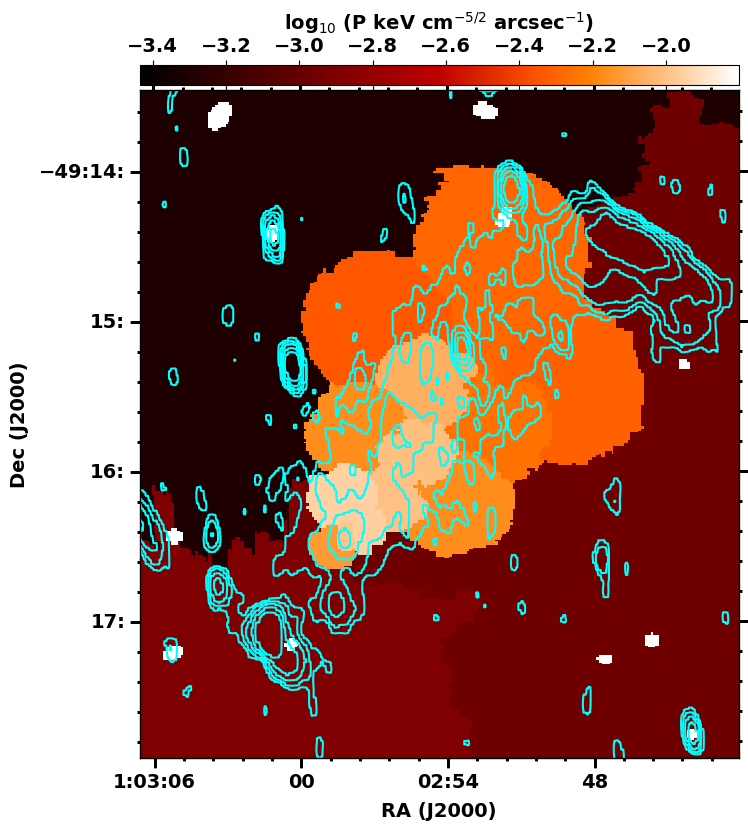}
\includegraphics[height=8.5cm]{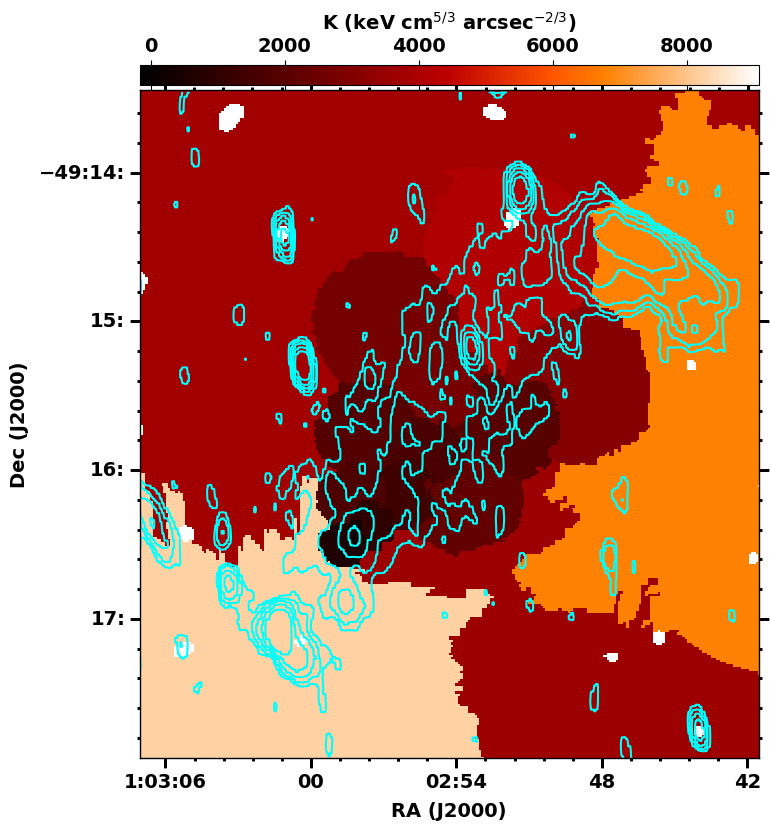}
\caption{\textit{Top left:} The X-ray temperature map is shown in color. The blue dashed lines indicate the two X-ray tails also shown earlier in Fig.~\ref{eg-b3-fig1}. \textit{Top right:} The map showing percentage error on the temperature (kT) is shown in color. Errors are dominated by the uncertainty on the temperature determination, which is $\sim$ 5 \% in the core region and $\sim$ 10\% in the rest of the cluster while for the larger bins in the map, the temperature is not constrained. \textit{Bottom left:} The pseudo-pressure map in logarithmic units is shown in color. \textit{Bottom right:} The entropy (K) map is shown in color. The 672 MHz contours (cyan) from Figure~\ref{eg-b4b5} are overlaid in all the panels.}
\label{xraytemp}
\end{figure*}

We have also compared the distribution of diffuse radio emission with the X-ray temperature
(Figure~\ref{xraytemp}). The temperature of the two X-ray tails is very high, peaking
at 17.8 keV for the N tail and at 14.7 keV for the S tail, as measured from the spectral bins
along the two tails depicted in the top left panel of Figure~\ref{xraytemp}. We note, however,
that due to the very low effective area of \textit{Chandra} at high energy, the measurement of
temperatures $>$10 keV is critical, and it is impossible to confidently disentangle temperature
differences between adjacent ``hot'' bins or between the two tails, while the error from the
spectral fitting do not entirely reflect the real uncertainty range for the same reason.
Nevertheless, in the X-ray thermodynamic properties, we find a possible trend along the elongation of the X-ray surface brightness from the southeast to the northwest, with the pressure being higher and the entropy lower in the southeast and vice-versa in the northwest. The radio halo emission is co-spatial with the region showing this trend (Figure~\ref{xraytemp}).

\subsection{Radio halo and the density fluctuation power spectrum}\label{sec:powerspec}

\begin{figure*}
    \centering
    \hbox{\includegraphics[width=0.5\textwidth]{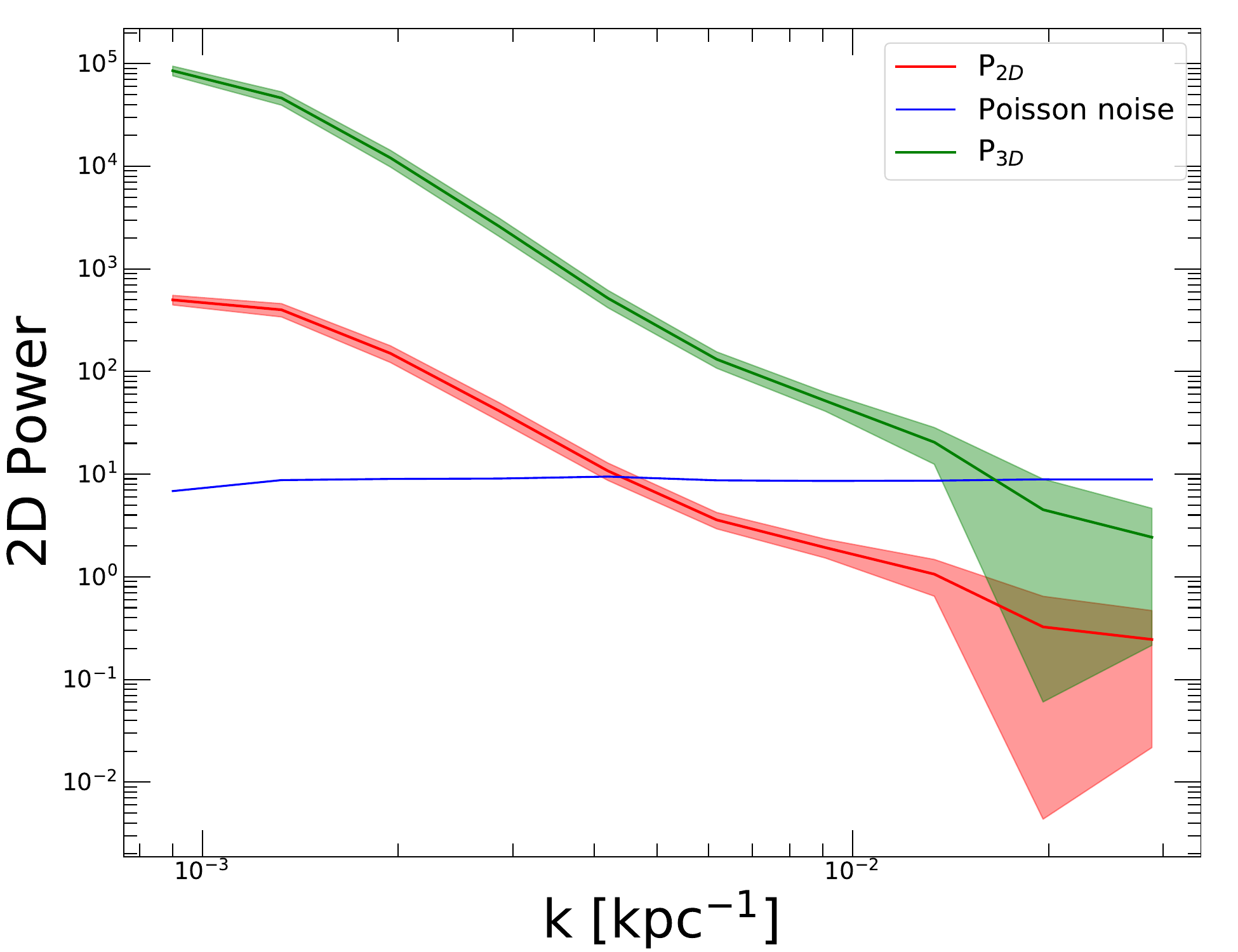}
   \includegraphics[width=0.5\textwidth]{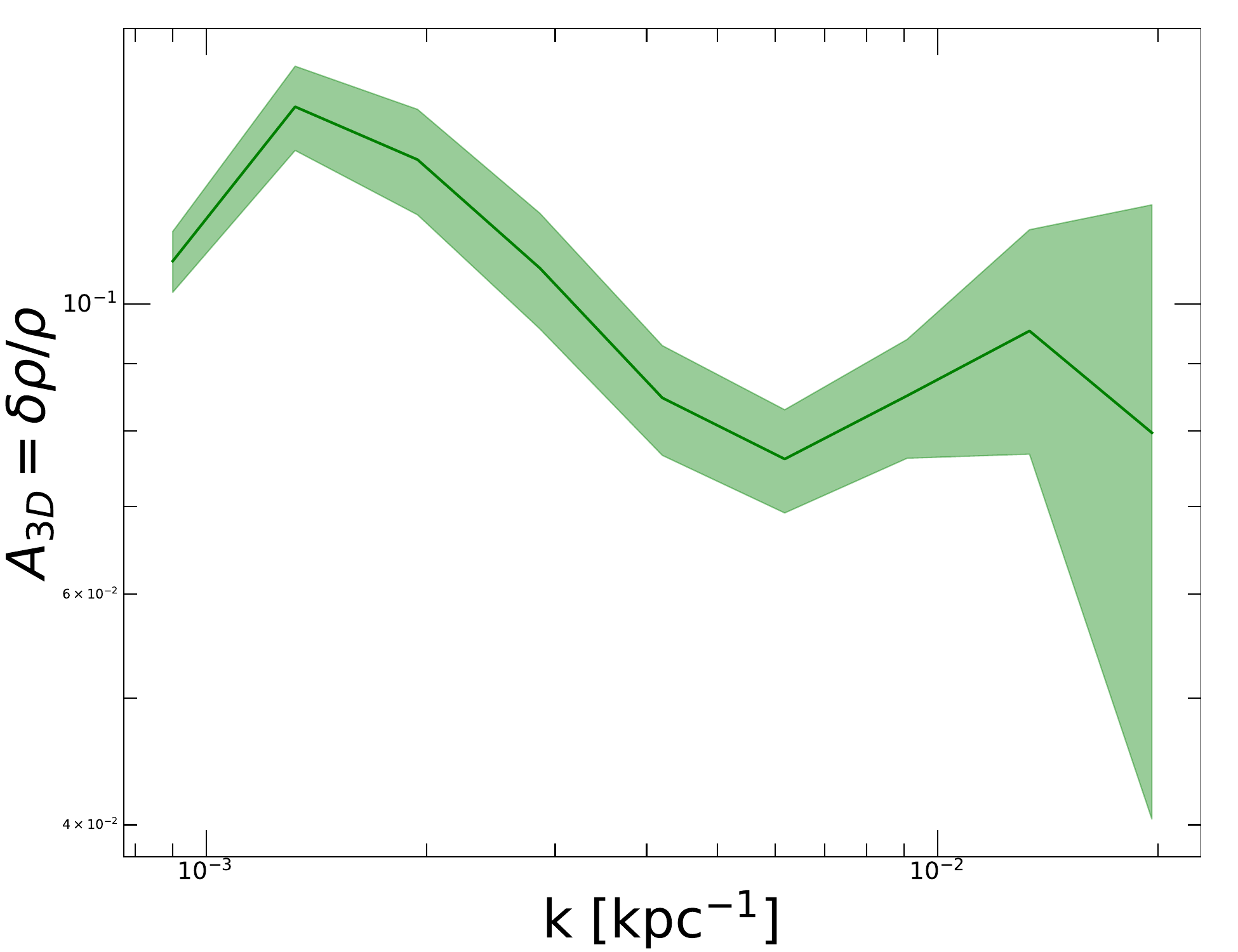}}
    \caption{\emph{Left:} Fluctuation power spectrum in the central 1 Mpc of El Gordo, where the ``bullet'' has been masked, as a function of wave number $k$. The red curve shows the 2D fluctuations in the power spectrum subtracted for the Poisson noise (blue curve). The green curve shows the corresponding deprojected 3D power spectrum. \emph{Right:} Amplitude of 3D fluctuations $A_{3D}=\frac{\delta\rho}{\rho}$ as a function of wave number.}
    \label{powerspec}
\end{figure*}

The power spectrum of density fluctuations derived using X-ray images can be used as a tracer of turbulent motions in the ICM \citep[e.g,][]{2012MNRAS.421.1123C,2013A&A...559A..78G,2015MNRAS.450.4184Z, 2023A&A...672A..42Z, Dupourque24}. Using the amplitude of density fluctuations at fixed spatial scales as a proxy for the level of turbulence, a scaling relation between the radio halo power and the velocity dispersion has been found
\citep{2017ApJ...843L..29E}. Here we applied a similar analysis to that performed in \citet{2017ApJ...843L..29E} to the \emph{Chandra} data of El Gordo. After masking the bright ``bullet'' structure around the X-ray peak (see Figure.\ref{eg-b3-fig1}), we used principal component analysis to determine the position of the centroid of the cluster and its ellipticity. We then extracted a surface brightness profile in elliptical annuli around the X-ray centroid, which we fitted in \texttt{pyproffit} \citep{eckert2020} with a single beta model \citep{cavaliere76}. From the best-fit brightness profile, we created a 2D elliptical model image accounting for CCD gaps and vignetting variations. Fluctuations on top of the model image were computed by dividing the true image by the model and filtering the residual image with 2D Mexican Hat functions, following the modified delta-variance method introduced by \citet{arevalo12}. The 2D power at a given scale is then proportional to the variance of the Mexican Hat filtered map. The 2D power spectrum of El Gordo within a circular region of 1 Mpc surrounding the X-ray centroid is shown in the left-hand panel of Figure~\ref{powerspec}, where we compare the resulting power spectrum with the contribution of Poisson noise.

Numerical simulations of galaxy clusters showed that the turbulent Mach number $\mathcal{M}=\sigma_v / c_s$ is linearly related to the maximum amplitude of gas density fluctuations, $A_{3D}=\frac{\delta\rho}{\rho}$ \citep{gaspari14,zhuravleva14}, where $\sigma_{v}$, $c_{s}$ are the velocity dispersion of the turbulent motion, and sound speed of the medium. To deproject the measured 2D power spectra and estimate the turbulent Mach number, we used the best-fit beta model to simulate a 3D galaxy cluster with properties similar to El Gordo, and we modified the 3D gas density field by a random 3D fluctuation field following a Kolmogorov power spectrum (it assumes a single injection scale). The perturbed 3D density field was then transformed into X-ray surface brightness and projected onto the line of sight to create a corresponding 2D surface brightness image. The ratio of 2D to 3D power spectra in the simulated data was then used to convert the 2D power spectrum into a 3D one, which can be seen as the green curve in the left-hand panel of Figure \ref{powerspec}. The 3D fractional amplitude of gas density fluctuations can then be retrieved as

\begin{equation}
    A_{3D}(k)\equiv \frac{\delta\rho}{\rho}=\sqrt{4\pi k^3 P_{3D}(k)/2}.
\end{equation}

with $k$ the wave number and $P_{3D}(k)$ the deprojected power spectrum. In the right-hand panel of Figure \ref{powerspec} we show the retrieved amplitude of gas density fluctuations in El Gordo as a function of wave number. The fluctuations peak at $L=k^{-1}\approx 700$ kpc at an amplitude $\max(\delta\rho/\rho)=0.15\pm0.02$. The approximate relation between density fluctuation amplitude and turbulent Mach number reads \citep{gaspari14}

\begin{equation}
    \mathcal{M}_{3D}\approx 4\max\left(\frac{\delta\rho}{\rho}\right)\left(\frac{L}{\rm 500 kpc}\right)^{0.25} \approx 0.63 \pm 0.06.
\end{equation}
For the ICM in El Gordo with $kT\sim14$ keV, the sound speed in the medium is $c_s = (\gamma kT /\mu m_p )^{1/2}\sim 1,915$ km~s$^{-1}$, such that the velocity dispersion of turbulent motions can be estimated as $\sigma_v=1201\pm88$ km~s$^{-1}$.

We caution the reader that the analysis presented here assumes that all the fluctuations in El Gordo should be attributed to turbulence. Although we masked the regions surrounding the bullet structure, inside which the fluctuations can be ascribed to a moving subhalo, there can remain additional sources of fluctuations on top of intracluster turbulence. Therefore, as for most complex merging clusters, the value retrieved here should be viewed as the maximum allowed value for the turbulent velocity dispersion.

\subsection{Radio halo power and cluster mass}

\begin{figure}
    \centering
        \includegraphics[width = \columnwidth]{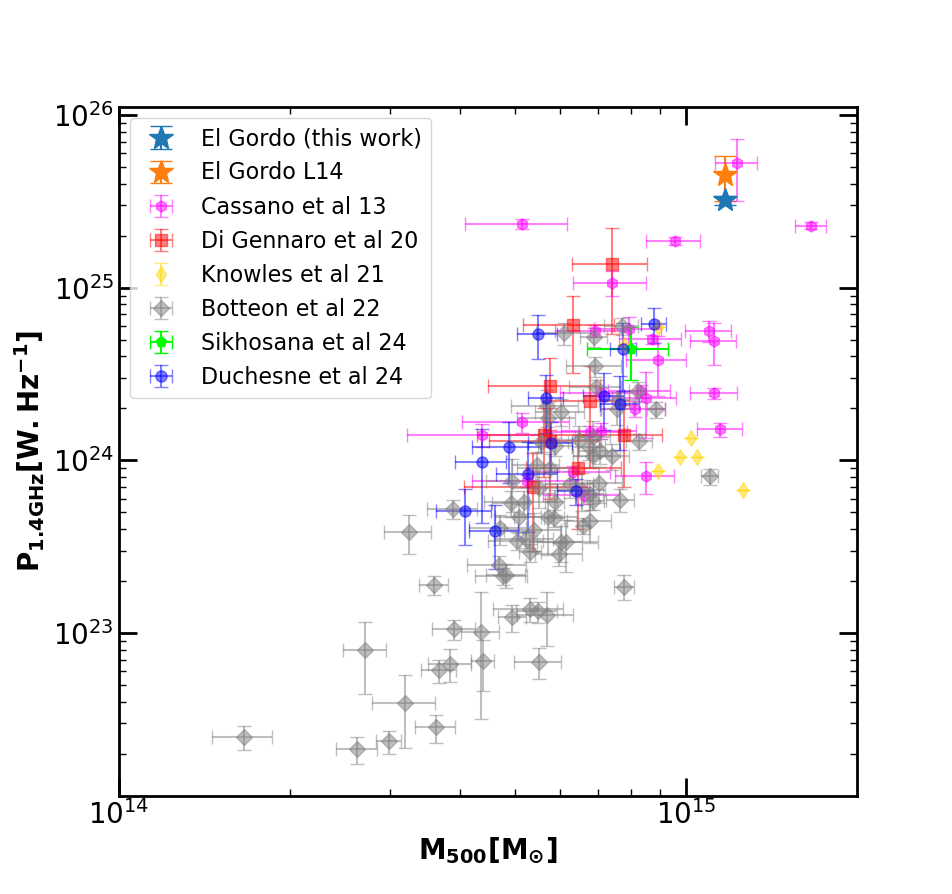}
    \caption{The 1.4 GHz radio power versus mass ($\rm{M}_{500}$) plotted for El Gordo (this work and L14) with other samples of radio halos from the literature.  
    The \citet{digen21} sample consists of the halos at redshifts \textgreater 0.6, the \citet{2021MNRAS.504.1749K} contains the radio halos in redshift range of 0.22 \textless z \textless 0.65 and the local redshift (z $\sim$ 0.2) sample from \citet{cas13}. Recent discovery of the radio halo in ACT-CL J0329.2-2330 (z $=1.23$)is also shown \citep{2024arXiv240403944S}.  Large samples from the LOFAR survey by \citet{2022A&A...660A..78B} ( 0.016 \textless z \textless 0.9), and from ASKAP by \citet{2024PASA...41...26D} (0.01 \textless z \textless 0.5) have been added to provide a more comprehensive picture.}
    \label{fig:p1400mass}
\end{figure}

The scaling relation between cluster masses and the radio power of radio halos and relics can be used to infer the connection between the non-thermal and thermal components of the ICM \citep{bas12,cas13, 2023A&A...680A..30C}. We have plotted the 1.4 GHz radio power  of El Gordo in the $\rm{P}_{1.4\rm{GHz}} - \rm{M}_{500}$ plane along with other samples from the literature (Figure~\ref{fig:p1400mass}). 
The L14 1.4 GHz value for El Gordo that was extrapolated from their 2.1 GHz measurement is also plotted for comparison. The current measurement and that of L14 are consistent within the uncertainties. El Gordo is the most radio-luminous among the high redshift samples \citep{digen21, 2024arXiv240403944S}. It is also the most massive among the high redshift sample and the radio power is in line with the expectation from the scaling relation. The radio powers for low-redshift clusters, reported at 150 MHz by \citet{2022A&A...660A..78B} and at 947 MHz by \citet{2024PASA...41...26D} are extrapolated to 1.4 GHz using a spectral index of $-1.3$, also shown in the same plot.

\section{Discussion}\label{discussion}

El Gordo is a high redshift powerful radio halo and double relic system. In the recent high redshift sample presented by \citet{digen21} there is only one relic: in PSZ2 G091.83+26.11 at $z=0.822$. A further candidate relic at high-redshift was reported in PSZ2 G069.39+68.05 ($z=0.762$) by \citealt{2022A&A...660A..78B}. This makes El Gordo so far the only cluster with a double radio relic system and a radio halo known at redshift $>0.6$. It is a peculiar system also in the X-ray band, where it shows the presence of a comet-like morphology with two tails. In this work, we presented the images and spatially resolved spectra of the radio halo and the relics using the uGMRT between 300 - 1450 MHz. We have detected the radio relics and the halo that were known earlier and fainter extensions of these are also found. The eastern relic has an extension further north making up a total extent of 1030 kpc ($134''$). There is a hint of detection of this feature also at band 4 but not at band 5 (Figure ~\ref{eg-b4b5}). The NW-relic is found to have the same spatial and spectral features as found by L14 and the integrated spectrum is a power-law with a spectral index of $-1.3\pm0.4$ and thus the physical picture of the origin of the radio relic in a shock of Mach number $\gtrsim3$ presented in \citet{2016MNRAS.463.1534B} still holds.

The radio halo is co-spatial with the X-ray emission as known earlier but the emission also extends over the southern tail (Figure ~\ref{eg-b3-fig1}). In the 610 MHz image using the legacy GMRT, it was clear that the radio halo preferred the northern X-ray tail. In deeper low-frequency observations at 366 MHz, the fainter parts of the radio halo that also extend over the southern X-ray tail are detected (Figure ~\ref{eg-b3-fig1}).  The integrated spectrum of the radio halo in El Gordo is a power-law with a spectral index $-1.0\pm0.3$ in the frequency range of 366 - 1274 MHz. However, between 1274 and 2100 MHz a steepening to $-2.50 \pm 0.46$ is observed (Figure ~\ref{intspec}). Curved radio spectra in radio halos have been found only in handful of clusters, Coma \citep{2003A&A...397...53T, murgia24}, AS1063 \citep{2020A&A...636A...3X}, MACSJ0717.5+3745 \citep{2021A&A...646A.135R}, A3562 \citep{vent22}. We emphasize that the 2100 MHz flux measurements may be biased due to the different \textit{uv-}coverage, and also the different regions have been used to estimate the flux density, however, more sensitive observations at 2000 MHz, and measurements with matched \textit{uv} range can shed light on the robustness of the steepening.

The origin of radio halos has been proposed to be in turbulent re-acceleration process \citep[e.g.,][]{bru01,pet01,cas05,bru07,bru&laz11,miniati15,bru16, nishiwaki&asano22}. It is a stochastic process that re-accelerates a mildly relativistic population of electrons. Hadronic collisions are present in the ICM and produce relativistic electrons, however, in many cases it has been shown that they are not sufficient to explain radio halos and the constraints from gamma rays \citep[e.g.,][]{bru08, bru17, wilber18, 2021A&A...648A..60A,2021A&A...650A..44B, 2021NatAs...5..268D, 2024arXiv240509384O, pasini24, nishiwaki24}. Turbulence and shocks are known to be injected during cluster mergers as shown by simulations and also observed in a number of clusters \citep[e.g.,][]{2000ApJ...542..608M,miniati15,2017MNRAS.464..210V}.

The point-to-point comparison between radio and X-ray images has been used as a tool to test the theoretical models \citep{2001A&A...369..441G}. The rationale behind the tool is that the gravitational potential energy from cluster mergers will dissipate as shocks and turbulence in the ICM - a fraction of that going into magnetic field amplification and particle acceleration. Under the assumption that the fraction of energy channeled into non-thermal components is independent of the position in the cluster, predictions for the radio-Xray surface brightness scaling can be obtained. The bolometric X-ray emissivity ($j_X$) is given by,

\begin{equation}
j_X \propto n_e^2 (kT_e)^{1/2},    
\end{equation}

where $n_e$ is electron density and $kT_e$ is the temperature. The thermal energy density ($\epsilon_{th} \propto 3 n_e KT_e$) squared equals the emissivity at a given position if the cluster is isothermal, resulting in,

\begin{equation}
j_X \propto \epsilon_{th}^2 (kT_e)^{-3/2}.    
\end{equation}

For the relativistic electrons, the number density between energies $\epsilon$ and $\epsilon + d\epsilon$, can be assumed to be $N(\epsilon) d\epsilon N_0 \epsilon^{-\delta} d\epsilon$. The radio emissivity is then given by,

\begin{equation}
    j_{Radio} \propto N_0 B^{(\delta+1)/2} \nu^{-(\delta-1)/2},
\end{equation}

where B is the magnetic field. For a linear relation between thermal and non-thermal plasma, it is expected that,

\begin{equation}
    N_0 B^{(\delta+1)/2} \propto n_e^2 (kT_e)^{1/2}.
\end{equation}

Under the assumption that the cosmic ray proton energy density ($\epsilon_{CRp}$) and magnetic field ($\epsilon_{B} = B^2/8\pi$) energy density is proportional to the thermal energy density ($\epsilon_{CRp} \propto \epsilon_{th}$ and $\epsilon_{B} = \epsilon_{th}$), for hadronic models 
the scaling relation is expected to be linear in the case of strong magnetic fields (as compared to the equivalent magnetic field for CMB) and super-linear in weak magnetic field assuming a radially decreasing magnetic field strength in the cluster. Super-linear scaling has been found in mini-halos  \citep[e.g.][]{2020A&A...640A..37I, ignesti22, Riseley:2023oge, lusetti24, biava24}. The turbulent re-acceleration can generate different slopes from super-linear to sub-linear or linear, depending on the parameters for the faint radio halos. If electrons (energy density) and magnetic fields (energy density) follow the thermal energy density, in the case of isothermal distribution and constant turbulent Mach number, the models produce the slope to be sublinear to linear (from strong to weak field) \citep[see eq.12][]{2024A&A...686A...5B}, as seen in the clusters A 2744 and A2255 \citep{gov01,2020ApJ...897...93B} and for 3 GHz scaling relation for the radio halo in MACS0717.5+3745 \citep{2021A&A...646A.135R}.

For El Gordo we found the relations to be sub-linear ($0.60\pm0.12$ at band 3 and $0.76\pm0.12$ at band 4). \citet{2023A&A...675A..51D} had reported a linear to super-linear correlation slope from 0.14-3.0 GHz for PSZ2 G091.83+26.11, at a redshift of 0.822. Other radio halos such as Coma \citep{gov01,2022ApJ...933..218B}, A520 \citep{2019A&A...622A..20H}, MACS J1149.5+2223 \citep{2021A&A...650A..44B} and AS1063 \citep{2020A&A...636A...3X} have sub-linear correlations. For MACS J0717.5+3745 at 144 and 1500 MHz \citep{2021A&A...646A.135R} the correlations are sub-linear but get closer to linear at higher frequencies. In El Gordo as well the band 4 (672 MHz) correlation slope is marginally higher than that at 366 MHz, though no firm trend can be established given the uncertainties. Such steeping can be attributed to the spectral steepening (the radial decay of the non-thermal plasma is different in frequencies) of the radio halo. Also, the SZ decrement at such a high redshift can play a role in the steepening. Further study utilising MeerKAT data at 1.28 GHz and other higher-frequency observations (above 1.5 GHz) will be needed to confirm this.

In the arguments regarding the expected correlation slopes, there are several assumptions such as isothermality of the cluster and that the fraction of energy that goes into the non-thermal components is independent of position in the cluster. From the thermodynamic maps of El Gordo, there is considerable variation in the properties across the cluster (Figure ~\ref{xraytemp}). The southern part of the radio halo, cospatial with the X-ray core has lower temperatures ($6 - 8$ keV) and higher pseudo pressure while the northern part has higher temperatures and lower pseudo pressure. We have also tried to investigate if the parts of the radio halo along the two X-ray tails behave differently in the $I_R - I_X$ correlation. However, with the current data, we are limited to only a few points (Figure ~\ref{ptp}), and thus it is not possible to find any trend. Future work combining MeerKAT data with deeper uGMRT observations could make it possible.

The spectral index is anti-correlated with the X-ray surface brightness with a slope of $-0.57\pm0.18$, and a correlation coefficient of $-$0.58. To date, the studies for the correlation between the spectral index and the X-ray surface brightness have been done for a handful of halos like A2255 \citep{2020ApJ...897...93B} and MACSJ0717.5+3745 \citep{rajp21}, A2256 \citep{rajp23}, A521 \citep{santra24}, A2142 \citep{riseley24}. An observed anti-correlation suggests spectral steepening in the outermost regions of the halo, indicating the faint X-ray emission probes the steep spectrum regions at the cluster outskirts. Since the radio halo is not circularly symmetric, the radial profile analysis will be hard to perform. However, this trend along with the increasing correlation slope implies the spectral steepening at the outskirts, similar to what is seen in L.14.

Using the X-ray fluctuation power spectrum approach described in Sec. \ref{sec:powerspec}, \citet{2017ApJ...843L..29E} derived an empirical scaling relation between the radio power of radio halos ($P_{1.4\rm{GHz}}$) and the three-dimensional turbulent velocity dispersion ($\sigma_v$), 

\begin{equation} \label{eq:8}
\log \left(\frac{P_{\rm 1.4 GHz}}{\rm 10^{24}\, W\, Hz^{-1}}\right) = \log P_{0} + \alpha \log\left(\frac{\sigma_v}{\rm 500\, km\, s^{-1}} \right).
\end{equation}
with $\alpha=3.27_{-0.61}^{+0.71}$, $P_{0}=2.34_{-0.49}^{+0.53}$, with an intrinsic scatter $\sigma_{\log P|\sigma_v}=0.44_{-0.13}^{+0.18}$. 
Given the total radio power of the radio halo in El Gordo, the scaling relation predicts $\sigma_v$ of 1112$_{-167.82}^{+230.06}$ km s$^{-1}$ (estimated using the error propagation). This value is close to and is consistent with, the value $\sigma_v = 1,201 \pm 88$ km s$^{-1}$ estimated from the amplitude of fluctuations in the ICM of El Gordo (see Sec. \ref{sec:powerspec}). The direct measurements of these motions will be available in future observations with XRISM. The equation~\ref{eq:8} is an empirical relation and is based on low-z radio halos, therefore, the obtained velocity dispersion can be considered as the lower limit. The X-ray and radio properties of the radio halo in El Gordo are consistent with the properties extracted from the GMRT radio halo sample \citep{2017ApJ...843L..29E}. The very high-velocity dispersion and radio power place the system at the high-power end of the relation, with only MACS J0717.5+3745 and the Bullet cluster (radio powers of (52.48 $\pm$ 20.56) $\times$ 10$^{24}$, (23.44 $\pm$ 1.51) $\times$ 10$^{24}$ W Hz$^{-1}$, and velocity dispersions of 1206 $\pm$ 96, 977 $\pm$ 99 km s$^{-1}$, respectively) exhibiting similarly high values for the turbulent velocity and radio power.

The energy rate per unit volume related to turbulence can be estimated as \citep[Eq.2][]{2017ApJ...843L..29E},

\begin{eqnarray}\label{e:Pturb}
\nonumber
P_{\rm turb} \approx  9.8\times10^{-25} \left(\frac{\sigma_{v}}{\rm 500\, km\, s^{-1}}\right)^3\left(\frac{n_{\rm gas}}{10^{-2}\rm \,cm^{-3}}\right) \\
\left(\frac{L_{\rm inj}}{\rm 500\,kpc}\right)^{-1} {\rm erg\,s^{-1}\,cm^{-3}}
\end{eqnarray}

where, $L_{\rm inj}$ is the injection scale and $n_{\rm gas}$ is the thermal gas density. For $L_{\rm inj}$ of 500 kpc, we estimate $P_{\rm turb}$ to be $1.07\times10^{-25}$ erg~s$^{-1}$~cm$^{-3}$. Simulations have proposed impact parameters of 300 kpc \citep{2015ApJ...800...37M} and 800 kpc \citep[][]{2015ApJ...813..129Z} for mergers matching the features of El Gordo. Using 800 and 300 kpc (assuming the scale of injection to be similar to the impact parameter) as the injection scales we obtain $P_{\rm turb}= 6.74\times10^{-26}$ and $1.79\times10^{-25}$ erg~s$^{-1}$~cm$^{-3}$, respectively.

The re-acceleration mechanism that generated the relativistic electrons in the radio halo has to work at a rate that balances the radiative losses over long timescales, and $\tau_{\rm max}$ is the maximum lifetime of the electrons that must be balanced by re-acceleration, given by (Eq. 14, \citet{bru16})

\begin{equation} \label{eq:10}
 \tau_{\rm max} = 156 \left(
{{\xi }\over{\nu_{o({\rm GHz})}}} \right)^{1/2}
(1+z)^{- 7/2} \,\,\,\,\, {\rm (Myr)}
\end{equation}

where $\nu_{o({\rm GHz})}$ is the observing frequency and $\xi$ is a factor in the range 6 $-$ 8 (we have used a value of 7) such that the steepening frequency, $\nu_s\sim \xi \nu_b$, where $\nu_b$ is the critical synchrotron frequency emitted by electrons having acceleration time same as their radiative lifetime \citep{cas05}. The magnetic field considered is given by, $B\sim B_{\rm CMB}/\sqrt{3}$ (the condition for maximum radiative lifetime). The observed frequency is $\nu_o = \nu_s/(1+z)$. The steepening frequency for the radio halo in El Gordo is likely to be between 1.274 and 2.1 GHz (Figure ~\ref{intspec}). Assuming steepening frequencies of 1.274 GHz and 2.1 GHz, we find $\tau_{\rm max}$ to be 41 Myr, and 31 Myr respectively. This timescale for the Coma cluster is $\sim 400$ Myr \citep[e.g.,][]{bru16}. Thus at the high redshift of El Gordo, the maximum available acceleration time is a factor of 10 shorter. The available acceleration timescale $\tau_{\rm acc}$ has to be shorter than the $\tau_{\rm max}$.

If we assume transit-time-damping (TTD) in the collisional regime, a reference model for radio halos \citep[e.g.,][]{bru07}, the acceleration time will be $\sim$ 70 $-$ 80 Myr (using the Mach number, sound speed, and injection scale), very close to what has been obtained using Equation~\ref{eq:10}. However, it should be noted that the $\tau_{\rm max}$ and the turbulent Mach numbers are upper limits, hence there is a hint that the efficiency needed for the acceleration is higher. This might suggest a situation where the effective mfp of ICM particles is reduced with respect to that due to Coulomb collisions, in this situation the acceleration time will be decreased, and the acceleration efficiency will be increased \citep[e.g.,][]{bru&laz11}. There is possible evidence for the reduced mean free path in the ICM \citep[e.g,][]{2019NatAs...3..832Z,digen21}. If we assume the non-resonant acceleration proposed in \citet{2016MNRAS.458.2584B}, the re-acceleration timescale will be $\tau_{\rm acc} \sim 450 \left(\psi/0.5 \right)^{0.3} {\rm (Myr)}$, (where $\psi = l_{mfp}/l_{A}$, and $l_{mfp}, l_{A}$ are the mean free path and Alfen scale respectively), implies a constraint on the mean free path of the relativistic electrons to $\leq$ 0.3, hence $l_{mfp} \leq l_{A}$. This is slightly smaller, compared to the reference value ($\sim$ 0.5) that has been motivated and used in the previous studies \citep[e.g.,][]{2016MNRAS.458.2584B, brun&vaz20}. Therefore, these extreme conditions may be used to check models and start suggesting that the efficiency constrained by observations is slightly higher than that expected by models in the standard/original configuration of model parameters.

\section{Summary and conclusions}

We have presented a study of the high redshift galaxy cluster El Gordo at 300 $-$ 1450 MHz using the uGMRT. The sensitive uGMRT observations allow us to study the detailed spectral characteristics of the radio halo below 1 GHz for the first time. Our observations, combined with available X-ray observations (\textit{Chandra}), provide acute physical insights into the thermal and non-thermal connections in the ICM, as well as the origin of the radio halo. We summarise our results and conclusions below:

\begin{enumerate}
    \item The uGMRT images of El Gordo at the effective frequencies of 366, 672, and 1274 MHz with rms noise of $50, 13$ and $15 \mu$ Jy beam$^{-1}$, respectively were obtained. We detected the radio halo, the NW Relic, SE Relic, and E Relic that were known from earlier studies. In addition, we detected an extension of the E Relic to the north that we have labeled ''E-relic Ext''.
    
    \item The radio halo emission is well fitted with a single power law between 300 $-$ 1274 MHz, with a spectral index of $-1.0\pm0.3$. The 2.1 GHz measurement from L14 indicates steepening beyond 1274 MHz, with a spectral index of  $-$2.50 $\pm$ 0.46; though the differences in the uv-coverage could affect this estimate. The integrated spectral indices of the NW Relic, SE Relic, and E Relic are also fitted with a single power law, with a slope of $-1.4$. The E-relic Ext has a steep spectrum of $-2.1$. The radio power of the halo at 1.4 GHz follows the P$_{1.4 \rm GHz}$ vs. M$_{500}$ relation for the high redshift halos and is the second brightest radio halo among all the radio halos across redshifts. 
    
    \item We obtained a spectral index map between 366 and 672 MHz which shows that the radio halo has a median spectral index of $-$1.3 and it shows a standard deviation of 0.4. However, the southern part shows a flatter spectral indices (\textgreater $-1.0$) as compared to the northern part where the spectral indices reach steeper values (\textless $-2$), suggesting a more complex situation with respect to a simple power law. The NW-relic also shows a gradient of the spectral index from the outer to inner edge.

   \item The brighter portion of the radio halo follows the northern X-ray tail. The Chandra temperature measurements at the edges of the radio emission are not robust enough to conclude about the co-spatiality with higher temperature regions and the brighter radio emission.
    
    \item The point-to-point comparison of 366 and 672 MHz images with the X-ray surface brightness show sub-linear fits, albeit with scatter at low surface brightness values. The relation was examined separately for the northern and southern X-ray tails. Both the regions are within the range where there is a large scatter in the scaling relation and the southern tail has only three points. The point-to-point comparison of the spectral index with the X-ray surface brightness shows that they are anti-correlated, indicating the brighter portions in X-ray have flatter spectra and vice versa.

    \item Using the radio power of the halo, we calculated the turbulent velocity dispersion based on the scaling by \citet{2017ApJ...843L..29E} to be 1112 km s$^{-1}$, implying a 3D turbulent Mach number of $\sim0.6$. The power spectrum of X-ray surface brightness fluctuations in the cluster provides a consistent estimate of the turbulent Mach number. Further assuming the turbulence injection scale to be the same as the cluster merger impact parameters proposed in simulations, we found the energy rate per unit volume of turbulence, $P_{\rm turb}$ to be $1.79\times10^{-25}$ and $6.74\times10^{-26}$ erg s$^{-1}$ cm$^{-3}$, if the typical scale of injection of turbulence is in the range 300 $-$ 800 kpc.

\end{enumerate}


\begin{acknowledgements}
The authors thank the referees for their comments that improved the paper.
  R.K. and R.S. acknowledge the support of the Department of Atomic Energy, Government of India, under project no. 12-R\&D-TFR-5.02-0700. R.K. also acknowledges the support from the SERB Women Excellence Award WEA/2021/000008. We thank the staff of the GMRT that made these observations possible. GMRT is run by the National Centre for Radio Astrophysics of the Tata Institute of Fundamental Research. This research has made use of the data from the GMRT Archive. This research has made use of data obtained through the High Energy Astrophysics Science Archive Research Center Online Service, provided by the NASA/Goddard Space Flight Center. This research has made use of  NASA's  Astrophysics Data  System, and of the  NASA/IPAC  Extragalactic Database (NED) which is operated by the Jet  Propulsion Laboratory, California Institute of Technology, under contract with the National Aeronautics and Space Administration.
\end{acknowledgements}

%
%
\bibliography{sample631}{}
\bibliographystyle{aa}

\begin{appendix} 
\section{Regions used for flux density measurements} \label{diff-regions}
The regions used for the flux density measurements of the diffuse sources described in Sec.~\ref{intspectra} are shown.

\begin{figure} [h!] 
\centering
      \includegraphics[height=7.6cm]{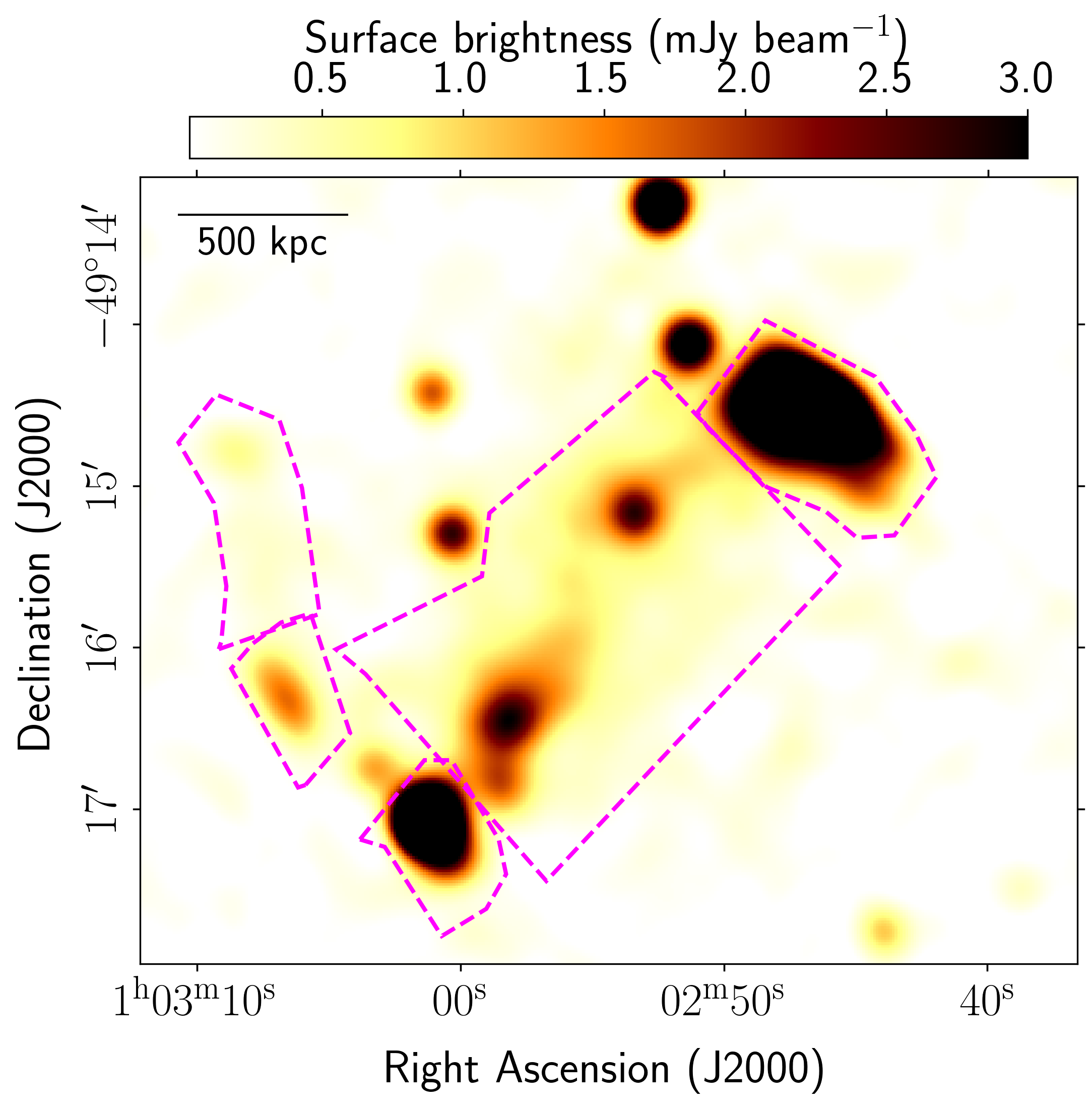}    
      \caption{The uGMRT band 3 image with a resolution of $16''\times16''$ is shown in color with the regions used to measure the flux densities of the diffuse sources overlaid in dashed lines.}
    \label{flux-regions}
\end{figure}

\section{Grids used for the point to point analysis} \label{grid}

We show here the regions used to extract the radio and X-ray surface brightness, corresponding to Figure.~\ref{ptp}.

\begin{figure} [h!] 
\centering
 
      \includegraphics[trim =0cm 1cm 2cm 2cm,clip,,height=7.5cm]{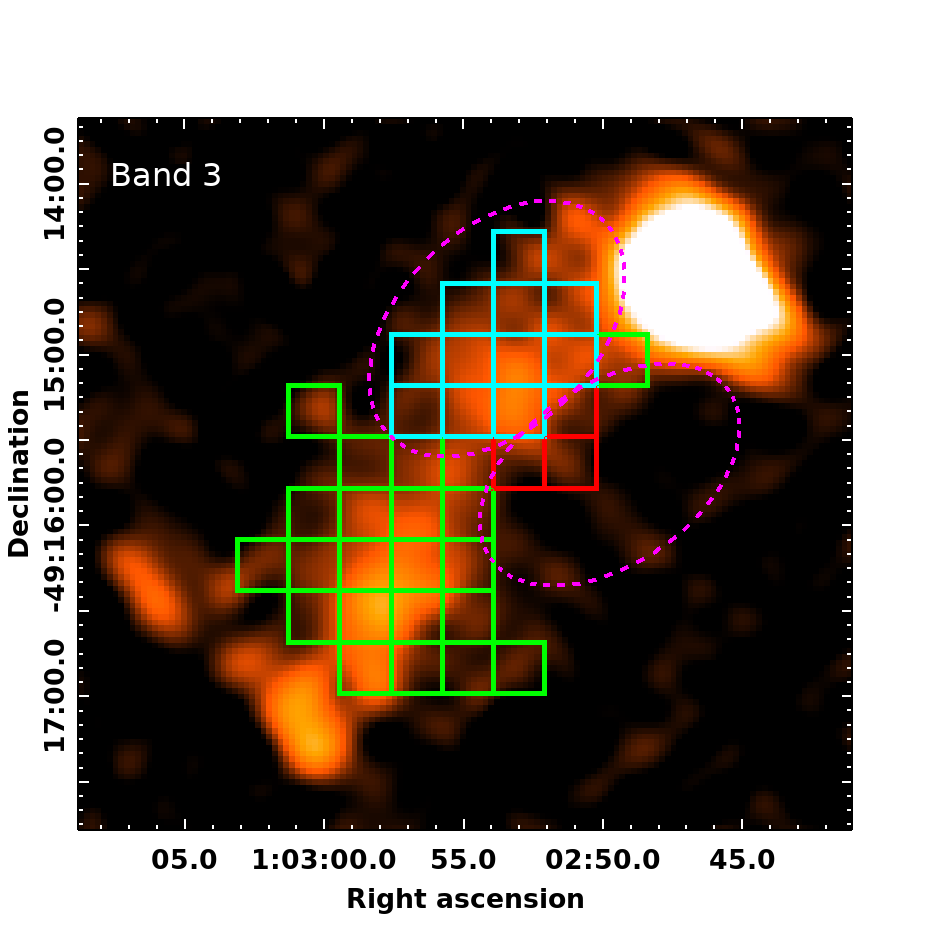}
      \includegraphics[trim =0cm 1cm 2cm 2cm,clip,,height=7.5cm]{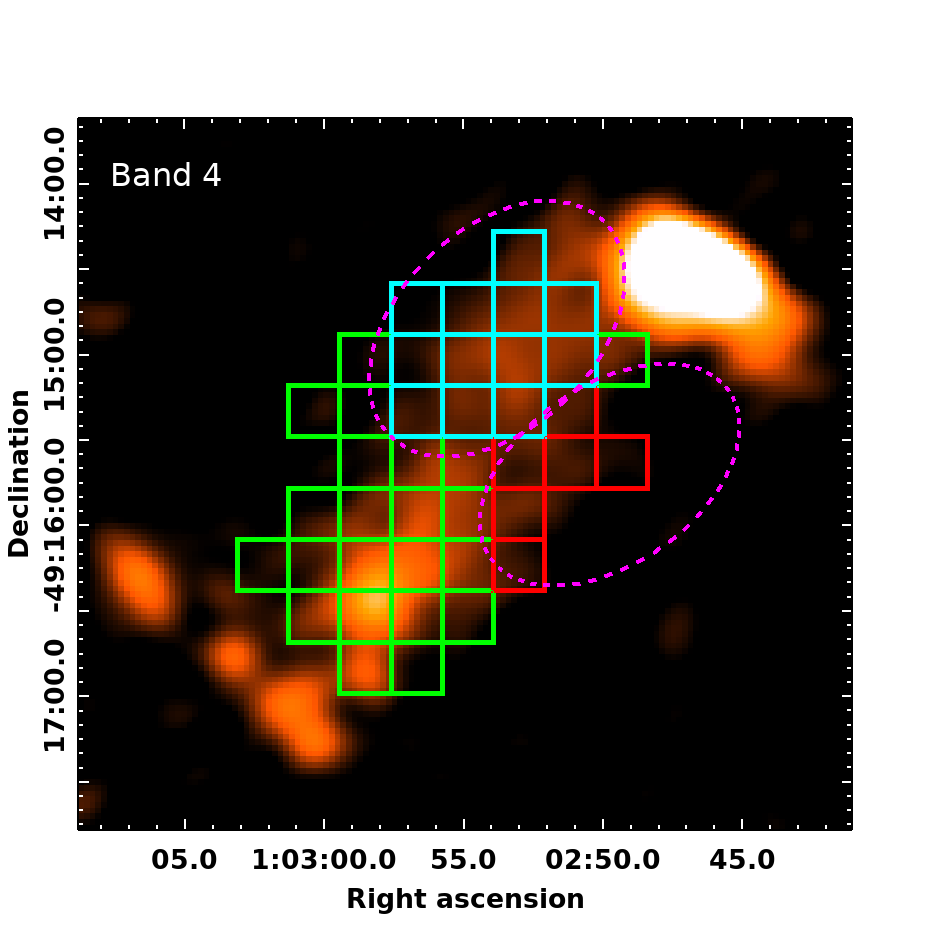}
      \caption{The two panels show the band 3 (top) and band 4 (bottom) images used in the spectral and point-to-point analysis. The two ellipses in magenta (dashed line) mark the two tails in X-rays. The regions of the grid within the two tails are color coded (cyan in the northern tail and red in the southern tail) in the point-to-point analysis. The rest of the regions are shown in green.}
    \label{ptpgrid}
\end{figure}

\FloatBarrier

\end{appendix}

\end{document}